\documentclass[aps,prx,onecolumn,nofootinbib,superscriptaddress,notitlepage,10pt]{article}
\usepackage{authblk}

\newcommand{\hi}{\mathcal{H}}

\newcommand{\his}{\mathcal{H}_{\mathcal{S}}}

\newcommand{\hir}{\mathcal{H}_{\mathcal{R}}}
\newcommand{\hirprim}{\mathcal{H}_{\mathcal{R}'}}
\newcommand{\hisr}{\mathcal{H}_{\mathcal{S}} \otimes \mathcal{H}_{\mathcal{R}}}

\newcommand{\Y}{\yen}

\newcommand{\Eff}{\mathcal{E}}

\newcommand{\R}{\mathcal{R}}
\renewcommand{\S}{\mathcal{S}}
\newcommand{\St}{\mathscr{D}}

\newcommand{\F}{\mathcal{F}}

\newcommand{\T}{\mathcal{T}}

\newcommand{\Cn}{\mathbb{C}}
\newcommand{\Reals}{\mathbb{R}}

\newcommand{\id}{\mathbb{1}}

\newcommand{\E}{\mathsf{E}}

\newcommand{\h}{\hspace{2pt}}
\newcommand{\ha}{\hspace{3pt}}
\newcommand{\haa}{\hspace{5pt}}

\usepackage{physics}
\usepackage{soul}
\usepackage{relsize}
\usepackage{lmodern}
\usepackage{slantsc}
\usepackage{bbm}
\usepackage{graphicx}
\usepackage{amsmath}
\usepackage{bbold}
\usepackage{amsfonts}
\usepackage{amssymb}
\usepackage{amsthm}
\usepackage{amsbsy}
\usepackage{mathrsfs}
\usepackage{varioref}
\usepackage{dsfont}
\usepackage{bm}
\usepackage{color}
\usepackage[nopar]{lipsum}
\usepackage{lipsum}
\usepackage{multicol}
\usepackage{tikz-cd}
\usepackage{color}
\usepackage[T1]{fontenc}

\newtheorem{theorem}{Theorem}[section] % 1st argument is your name for it
\newtheorem{definition}[theorem]{Definition}

\usepackage{amsmath}
\usepackage[makeroom]{cancel}

\usepackage{enumitem}

\usepackage[utf8]{inputenc} %to manage special characters
\usepackage{fancyhdr} %to customize the headers
\usepackage[lmargin=0.6in, rmargin=0.6in, tmargin=0.5in, bmargin=1in]{geometry} %sets the margins for the pages
\usepackage{url} %to include urls
\usepackage{listings} %include this if you want to include code in the thesis
\usepackage{amsmath,amssymb} %mathematical package
\usepackage{array, booktabs} %to make better tables
\usepackage{graphicx} %to include graphics
\usepackage{float} %to include floats
\usepackage{color} %edit e.g. text colors

\usepackage[backend = biber,
            style = chem-angew,
            date = long,     % Long: 24th Mar. 1997 | Short: 24/03/1997
            sorting = none,
            maxcitenames = 3,   % max names to include before et. al.
            ]{biblatex} %customize the look of your citations and bibliography
\usepackage{comment} %to be able to comment out sections in the .tex files

\usepackage{hyperref}
\usepackage{cleveref}

\usepackage{pifont}

\title{
\textbf{%Relational Quantum Relativity
Towards Relational Quantum Field Theory
}
}

\author{Jan G\l{}owacki\thanks{jan.glowacki.research@gmail.com}}
\affil{\vspace{7pt}\emph{Department of Computer Science, University of Oxford, UK}}
\affil{\vspace{-3pt}\emph{Basic Research Community for Physics, Leipzig, GERMANY}}

\date{}

\begin{document}

\maketitle

\begin{abstract}
This paper presents a research program aimed at establishing relational foundations for relativistic quantum physics. Although the formalism is still under development, we believe it has matured enough to be shared with the broader scientific community. Our approach seeks to integrate Quantum Field Theory on curved backgrounds and scenarios with indefinite causality. Building on concepts from the operational approach to Quantum Reference Frames, we extend these ideas significantly. Specifically, we initiate the development of a general integration theory for operator-valued functions (quantum fields) with respect to positive operator-valued measures (quantum frames). This allows us to define quantum frames within the context of arbitrary principal bundles, replacing group structures. By considering Lorentz principal bundles, we enable a relational treatment of quantum fields on arbitrarily curved spacetimes. A form of indefinite spatiotemporality arises from quantum states in the context of frame bundles. This offers novel perspectives on the problem of reconciling principles of generally relativistic and quantum physics and on modelling gravitational fields sourced by quantum systems.
\end{abstract}

\vspace{0.3in}

\begin{multicols}{2}
\section{Introduction}

Quantum Field Theory, understood in terms of quantized relativistic fields over the Minkowski spacetime and enriched with calculational techniques allowing for pictorial representations and plausible interpretation of interactions, provides a conceptual and formal foundation for the Standard Model. It is arguably the most successful empirical model of microscopic phenomena in physics. However, it is uncontroversial to point out that this framework lacks a particularly firm mathematical foundation, and the naive ontology it suggests is troublesome. So far, it has also fallen short of providing a means for reconciliation with modern theories of gravitation. Since this situation persists for many decades now, it seems appropriate to look for alternative ways in which relativistic principles can be incorporated into Quantum Theory.

Foundational research in QFT has delivered plausible axiomatisations and coherent mathematical formalisations for the idea of a quantum field (e.g \cite{streater_pct_2000}, \cite{haag_local_1996}), which however do not seem compatible with the sophisticated models of interacting theories and the ways they are used in physics to derive very accurate predictions.\footnote{Substantial progress on this problem has been made in the last years with a general framework for quantum measurement theory in QFT being developed (see e.g. \cite{fewster_measurement_2023} for a recent exposition), various models for coupling detectors to quantum fields investigated (e.g \cite{polo-gomez_detector-based_2022,perche_particle_2024}), and philosophical considerations of the related issues being carried out in the community (e.g. \cite{fraser_note_2023}). Incorporating the formalism presented in this work into modeling local yet fully relativistic detectors in terms of quantum reference frames is being investigated.} 

Here we take a different route and begin by introducing Special Relativity (SR) into Quantum Theory (QT) via a particular formalism for Quantum Reference Frames (QRFs), referred to as \emph{operational} \cite{carette_operational_2023,glowacki_operational_2023} (see also preceding developments \cite{loveridge_quantum_2012,loveridge_symmetry_2018,loveridge_relativity_2017,loveridge_relative_2019,loveridge_relational_2020}, and recent advances \cite{glowacki_relativization_2024,glowacki_quantum_2024,fewster_quantum_2024}). The framework focuses on the notion of relative observables that are invariant with respect to an underlying symmetry group and contingent upon the experimental arrangement employed in the measurement, reflected by the choice of a quantum frame. The general idea is then to \emph{relativize} QT and emphasize \emph{operationality} by introducing the Poincar{\'e} group as the underlying symmetry structure into the operational approach to QRFs. From this perspective, we recover local observables of QFT as relational quantities taken with respect to an experimental arrangement localized in a spacetime region.

These intuitions are then generalized to the realm of gauge theories, by the means of newly developed mathematical tools allowing for integration of a wide class of operator-valued functions with respect to arbitrary positive operator-valued measures (POVMs). Thus, a definition of a quantum reference frame on a \emph{principal bundle} is given -- as a \emph{covariant} POVM on a subset of the bundle lying above a space-time region, together with a \emph{local section} --  with the relativization procedure extended to the realm of quantum fields. The resulting relative operators are invariant under global gauge transformations, while locally the gauge is fixed by the choice of the section. The subsets of the base manifold (understood as regions of spacetime) are only meaningful as relational entities -- they represent the `place (and time)' where (and when) the quantum frame and the quantum field `interact'.

The notion of an \emph{external quantum reference frame transformation} is also extended to such QRFs, generalizing and unifying the notions of a local coordinate transformation, a change of gauge, and a change of the physical system used as a reference. We believe this may be the notion relevant for investigating covariance with respect to quantum frames. Another new notion - that of a \emph{reduction} of the symmetry structure of a QRF - is proposed end exploited. It describes how frames defined on sub-groups and sub-bundles can serve as references for systems subject to larger symmetry structures. The definition of relational local observables is then extended to the case of quantum fields subject to arbitrary gauge symmetries.

These advances allow us to meaningfully apply the framework to the context of curved geometries by capturing a Lorentzian manifold in terms of a principal bundle for the Lorentz group; quantum frames then provide local tetrad fields in a probabilistic way contingent on the quantum state. By invoking the frame bundle as the underlying symmetry structure we arrive at a setup encompassing \emph{non-inertial} quantum reference frames and \emph{indefinite background geometries}. We speculate how these newly discovered models may help reconcile the geometrical and relational nature of General Relativity (GR) with the formalism of QFT. We believe the introduced tools may also be useful in modelling gravitational fields of superposed massive objects (see e.g. \cite{chen_quantum_2024}).

The  relational framework for relativistic quantum fields emerging from our considerations reveals deep similarities and important differences with the established approaches to QFT. The concepts of quantum fields and smeared field observables arise naturally, although are represented differently as mathematical objects. The quantum fields arise as ultraweakly continuous and bounded \emph{operator-valued functions} (or equivalence classes thereof) on a space-time manifold; a suitable approach to partial differential equations to be satisfied by such objects is proposed. The smeared observables arise relationally as integrals of a quantum field taken with respect to \emph{probability measures} describing localization of a \emph{quantum reference frame}.  The exact relation of the approach to relativistic quantum physics proposed in this work and the established formulations of QFT is being investigated in separate work.

We mention here various other developments in the intersection of the quantum reference frames program with relativistic and gravitational physics. These include the formalisms revolving around the idea of `superposing geometries' (e.g.  \cite{kabel_identification_2024,giacomini_spacetime_2021}), alternative to the operational approach attempts to realize special-relativistic symmetries in the context of QRFs (e.g. \cite{de_la_hamette_perspective-neutral_2021}), and intersecting the operational approach to QRFs with quantum measurement theory in the context of QFT \cite{fewster_measurement_2023,fewster_quantum_2024}. We make a precise connection to this last-mentioned development in Sec. \ref{RQFTcurved}.

The formalism proposed here differs widely from all the mentioned developments, both in terms of its scope -- we propose a way to treat general quantum field theories on arbitrary backgrounds and in the context of indefinite causality, and in terms of the mathematical implementation -- we develop new mathematical tools to realize our ideas. This is achieved via a different approach to \emph{symmetries} in the context of spacetimes being employed: instead of considering the group of diffeomorphisms, which in most cases is mathematically intractable, or treating only highly symmetric background manifolds, which we find too restrictive, we employ the tools of the theory of principal bundles which we believe to be advantageous in these respects.

After the first version of this text was completed, a different approach to quantum reference frames on frame bundles was announced \cite{vanzella_frame-bundle_2024}. The main goal of that work was to make formal sense of the idea of superposing geometries to the extent it may be relevant to modelling gravitational fields sourced by a mass in a coherent superposition of position states. It is then remarkable how close the employed methods come to those presented here. In our current understanding, the notion of a quantum reference frame presented there corresponds to that of an ideal frame as defined here, together with a choice of a pure state. We touch upon this relation in the discussion, postponing the full comparison -- and perhaps convergence -- of the two approaches to future work.

\section{Relativistic Quantum Frames}\label{QRFs}

Our story begins with merging the basic principles of SR: that physical systems ought to be described with respect to reference frames and that all inertial frames are equally well-suited for this task, with the recently introduced mathematical realization of a notion of a relative state \cite{carette_operational_2023}. This way we introduce the operational QRF formalism into the realm of SR, which has not been attempted before in the literature. Two notions absent from the recent expositions of the framework \cite{carette_operational_2023,glowacki_operational_2023}: that of an \emph{external frame transformation} \cite{glowacki_relativization_2024}, and a new general procedure of \emph{reducing} the symmetry group of a frame, are introduced.

\subsection{Relativistic relative states}

In the Special Theory of Relativity, physical systems are described with respect to inertial frames. Thus, if the system in question is modelled within quantum theory, the quantum state capturing its probabilistic properties ought to be given only upon specification of an inertial frame. Consider the set of such frames -- denote it $F$. Any pair of frames $x,x' \in F$ needs to be related by a unique Poincar{\'e} transformation, which gives $F$ the structure of a \emph{torsor}, i.e, the Poincar{\'e} group $G$ acts on $F$ freely (with no fixed points) and transitively (any point can be reached from any other). As a result, $F$ is in bijection with $G$, and we will understand $F$ to be equipped with topological, differential and measurable structures inherited from $G$, as needed. Now consider a quantum state (density operator) of the system $\S$ relative to a chosen frame $x \in F$, written as $\rho^{(x)} \in \St(\his)$. According to SR, such a state should be given for any choice of frame, so that we have a map
\begin{equation}\label{eq:relst1}
     F \ni x \mapsto \rho^{(x)} \in \St(\his).
\end{equation}
Defining $g.\rho^{(x)} := \rho^{(g.x)}$ for $g \in G$ and allowing for arbitrary $\rho^{(x)}$ brings us to considering an action of $G$ on $\St(\his)$; if we expect this action to preserve pure states, it will be implemented by a unitary representation\footnote{In this paper, contrary to the previous literature on the subject, we will work with the convention that the action of $G$ on states is left and, as a consequence, the action on operators is right (written $A.g$ for $A \in B(\hi)$). This is to align ourselves with the generalization of the formalism the the realm of gauge theories where typically the action on the principal bundles is right.} $g.\rho = U(g)\rho U(g)^*$. The map \eqref{eq:relst1} is then \emph{equivariant} by definition. Using this minimal setup, we could say that before the frame has been chosen, the state of $\S$ is only specified up to Poincar{\'e} transformation; in other words, it is given by the \emph{orbit} $[\rho]_G \in \St(\his)$.

As a next step towards considering the frames as quantum systems too, consider now a classical system $\R$, think of it as a bunch of rulers or screens and a clock, that can provide us with a local coordinate system for our observations, say on the table in our lab. Such a coordinate system will necessarily be deficient in the sense of unavoidably limited precision of the measurement of distances and time differences it can provide. Worse, consider a situation when we can not be sure if the frame has been somehow rotated or otherwise transformed with respect to the system we will be measuring -- this is quite a realistic scenario in the case when other people are using our lab! Nevertheless, let's imagine we can assign a probability distribution to the actual orientation of the frame, e.g, we think it stayed as we left it the previous day, so that we are using $x \in F$, with probability $q \in [0,1]$, and that our colleague rearranged it by $g \in G$ to perform their measurements in $x'=g.x \in F$, with probability $(1-q)$. Such uncertainly could be implemented into our experiment by flipping a weighted coin and performing or not the transformation $g$ on the frame. In such a situation, we would reasonably assign the state
\begin{equation}
     q \h \rho^{(x)}+(1-q)\h\rho^{(x')}  \in \St(\his),
\end{equation}
to the system $\S$. In general, given an arbitrary probability distribution $\mu \in {\rm Prob}(F)$ describing the uncertainty of the frame orientation, the corresponding relative state would~be
\begin{equation}
    \rho^{(\mu)} := \int_F \rho^{(x)} d\mu(x).
\end{equation}
Let us point here to a subtlety arising in this context. As mentioned before, the space of frames is a $G$-torsor, the only difference between $F$ and $G$ being that $G$ has a specified, distinguished identity element $e \in G$ whereas $F$ does not. We may then pick a concrete realization of the homomorphism $F \cong G$ by picking $x_e \in F$ as an arbitrarily chosen `origin'. We then have
\begin{equation}
    \int_F \rho^{(x)} d\mu(x) = \int_G g.\rho^{(x_e)} d\mu(g.x_e).
\end{equation}
Choosing a different origin $x'_e = h.x_e$ will give the same result which is readily seen by performing the corresponding change of variables.\footnote{Notice here, that $\mu$ need \emph{not} be the invariant (Haar) measure to carry this reasoning.} The relative state $\rho^{(\mu)}$ is then insensitive to the choice of the homomorphism $F \cong G$. From now on, we will use this ambiguity and write
\begin{equation}
    \rho^{(\mu)} = \int_G g.\rho \h d\mu(g).
\end{equation}

Finally, imagine -- since this seems far from experimental practice -- that the probabilistic uncertainty of the frame orientation is due to the system $\R$ being quantum; in other words, consider $\mu$ arising as a Born probability measure
\begin{equation}
    \mu (X) \equiv \mu^{\E_\R}_\omega (X) := \tr[\omega \E_\R(X)],
\end{equation}
where $X \in {\rm Bor}(G)$ is a (Borel) subset of $G$ and we have introduced $\R$ as a quantum system modelled on $\hir$ in a state $\omega \in \St(\hir)$, together with a positive operator-valued measure\footnote{The definition of a positive operator-valued measure has been included in App. \ref{POVMs}.} (POVM)
\begin{equation}
    \E_\R: {\rm Bor}(G) \to \Eff(\hir).
\end{equation}
When $\E_\R$ has been specified, the quantum state of $\S$ can now be understood as given relative to a quantum state $\omega \in \St(\hir)$ which we write as \cite{carette_operational_2023}
\begin{equation}\label{weightedstate}
    \rho^{(\omega)} := \int_G g. \rho \h d\mu^{\E_\R}_\omega(g).
\end{equation}

The action of $G$ on $F \cong G$ naturally generates the action\footnote{As already noted, we work with the right action on operator algebras, which enters here the definition of covariance.} of $G$ on the effects of $E_\R$ by
\begin{equation}
    \E_\R(X).g := \E_\R(X.g),
\end{equation}
where $X.g = \{x.g \ha | \ha x \in X\}$. This makes $\E_\R$ a \emph{covariant} POVM. As before, since we would not like to restrict arbitrariness of $\E_\R$, this brings us to assuming an action of $G$ on all of $\Eff(\hir)$, which then uniquely extends to the operator algebra $B(\hir)$. In other words, a quantum frame (QRF) arising from our considerations is a quantum system with a $G$-action equipped with a covariant POVM on $G$, which will be referred to as the \emph{frame observable}.

\subsection{Relativization}

The concept of quantum states being given relative to a quantum reference frame in a specified state introduced above can be captured as a map taking a pair of states and assigning to them the corresponding relative state:
\begin{equation}
    \St(\his) \times \St(\hir) \ni (\rho, \omega) \mapsto \rho^{(\omega)} \in \St(\his).
\end{equation}
This map extends to a quantum channel
\begin{equation}
    \St(\hisr) \ni \rho \otimes \omega \mapsto  \rho^{(\omega)} \in \St(\his),
\end{equation}
which, due to the covariance of the frame observable underlying this assignment, is well-defined on orbits of the diagonal group action on the composite system in that  for any $h \in G$ we have \cite{carette_operational_2023}
\begin{equation}\label{relstclass}
    h.(\rho \otimes\omega) = h.\rho \otimes h.\omega \mapsto (h.\rho)^{h.(\omega)} = \rho^{(\omega)},
\end{equation}
where we have used $\rho^{h.(\omega)} = h^{-1}.\rho^{(\omega)}$ (which is easily confirmed, and to be found in \cite{carette_operational_2023}). The \emph{dual} of this channel is the \emph{relativization map} \cite{loveridge_quantum_2012,carette_operational_2023} defined as
\begin{equation}\label{eq:yen}
    \hspace{-2pt}\Y^\R: B(\his) \ni A \mapsto \int_G A.g \otimes d\E_\R(g) \in B(\hisr).
\end{equation}
Indeed, one easily verifies that
\begin{align}
\begin{split}
\tr[\rho^{(\omega)}A] 
&= \int_G \tr[g.\rho \h A] d\mu^{\E_\R}_\omega = \int_G \tr[\rho \h A.g] d\mu^{\E_\R}_\omega \\
&= \tr[\rho \otimes \omega \int_G A.g \otimes d\E_\R(g)],
\end{split}
\end{align}
where the last equality can be understood as the defining property of the operator $\Y^\R(A)$, as elaborated further below. As a consequence of covariance of $\E_\R$ the image of the relativization map lies in the invariant algebra $B(\hisr)^G$, which corresponds to \eqref{relstclass}. Indeed, for any $h \in G$ we have
\begin{align} \begin{split}
    \Y^\R(A).h &= \left(\int_G A.g \otimes d\E_\R(g)\right).h \\ &= \int_G A.gh \otimes d\E_\R(gh) = \Y^\R(A).
\end{split} \end{align}
The operators in the (ultraweak closure of the) image of $\Y^\R$ are called $\R$-relative, written $B(\his)^\R:= \Y^\R(\his)^{\rm cl}$. The relativization map is a quantum channel and thus can be used to generate relative observables (POVMs) from those on $\S$ alone through composition.

\subsection{Restriction}

The operational approach to QRFs explores the hypothesis that the observables accessible to the experimenter are relative in the sense of arising through the relativization map. The natural question arises how it then can be that the standard, non-relational description of $\S$  in terms of $B(\his)$ is working so well? The proposed answer is that in experimental practice, we use quantum reference frames with classical-like properties prepared in what we call localized states. To be concrete, a QRF is called \emph{localizable} if there is a sequence of states $\omega_n$ such that $\lim_{n \to \infty} \mu^{\E_\R}_{\omega_n} = \delta_e$. Now when the reference is prepared in fixed state $\omega \in \St(\hir)$, description of $\S$ relative to $\R$ is suitably conditioned by means of the \emph{restriction map}, which amounts to plugging in the given state of the frame
\begin{equation}
    \hspace{-4pt}\Gamma_\omega: B(\his)^\R \ni \Y^\R(A) \mapsto \int_G A.g \h d\mu^{\E_\R}_\omega(g) \in B(\his).
\end{equation}
We will write $\Y^\R_\omega = \Gamma_\omega \circ \Y^\R$ for the composition of relativization and restriction maps. Similar to what was happening with relative states \eqref{weightedstate}, the observable $A$ is weighted with the probability distribution for the orientation of the frame in a given state. Thus if the frame is localizable and prepared in a highly localized state $\omega_n$, the observable $A$ can be approximated by a conditioned relative observable $\Y^\R_{\omega_n}(A)$ to arbitrary precision \cite{carette_operational_2023}.

\subsection{Reduction}

We now introduce a simple although new\footnote{A special case of a restricted symmetry group setup was considered and discussed in \cite{loveridge_symmetry_2018}.} concept in the general theory of operational QRFs, that turns out to be of importance in our generalization to gauge theories and quantum fields in the sequel. To this end, imagine that a frame that we have under our disposal is not sensitive to all of the degrees of freedom of $G$ but can only `resolve' those associated with a subgroup $G_\R < G$. As an example, one can think of a spin or position measurement of a relativistic quantum particle. This corresponds to a frame observable being defined on a subgroup\footnote{We will use $\mathsf{F}_\R$ to denote the frame observables defined on the reduced symmetry structures, and reserve $\E_\R$ to those pertaining to the full symmetry structure of the system that is to be described in relation to $\R$.}
\begin{equation}
\mathsf{F}_\R: {\rm Bor}(G_\R) \to \Eff(\hir)
\end{equation}
of the group of symmetries of the system, e.g. Poincar{\'e} group in the special relativistic context considered here. Such scenario can be described in our framework by \emph{pushing forward} a covariant POVM on $G_\R$ to the one defined on $G$ via the inclusion $i_{G_\R}: G_\R \hookrightarrow G$, i.e,
\begin{equation}
    \E_\R := \mathsf{F}_\R \circ i_{G_\R}^{-1}: {\rm Bor}(G) \to \Eff(\hir).
\end{equation}
Such a frame observable will be $G_\R$-covariant, and give rise to relative $G_\R$-invariant operators. Indeed, we have\footnote{The change of variable for an integral taken with respect to a POVM will be justified shortly in Sec. \ref{ovi}. A version of this formula and its interpretation appeared in \cite{loveridge_symmetry_2018}.}
\begin{equation}
    \Y^\R(A) = \int_G A.g \otimes d(\mathsf{F}_\R \circ i_{G_\R}^{-1})(g) = \int_{G_\R} A.g \otimes d\mathsf{F}_\R(h).
\end{equation}
We refer to this procedure as \emph{reduction} of the group symmetry structure. It will be generalized to the context of principal bundles in Sec. \ref{sec:QFaF}.

\subsection{External frame transformations}

Finally, we mention here the concept of \emph{external} frame transformations introduced in \cite{glowacki_relativization_2024}. In this approach, a change of the description of a chosen system $\S$ given relative to a QRF $\R$ is transformed into a description of $\S$ given relative to another frame, $\R'$, \emph{along an equivariant channel} $\psi: B(\hir) \to B(\hirprim)$ such that
\begin{equation}
    \E_{\R'} = \psi \circ \E_\R.
\end{equation}
This is achieved by simply applying the $\id_\E \otimes \psi$ channel to observables relative to $\R$ since we have
\begin{align} \begin{split}
    \Y^{\R'}(A)
    &= \int_G A.g \otimes d\E_{\R'}(g) \\
    &= \int_G A.g \otimes d(\psi \circ \E_{\R})(g)\\
    &=(\id_\S \otimes \psi) \int_G A.g \otimes d\E_{\R}(g),
\end{split} \end{align}
(which uniquely extends to $B(\his)^\R$ by continuity). Such an equivariant channel then gives a map
\begin{equation}\label{eq:eft}
    \Y^\R(\psi): B(\his)^\R \to B(\his)^{\R'}.
\end{equation}
Frame transformations of this kind are called external since, contrary to all other proposals in the literature, no `global' perspective involving all three systems $\S$, $\R$ and $\R'$ is ever invoked. We see this as better aligned with the principles of operational quantum physics \cite{busch_quantum_1996}.

\subsection{Rediscovering quantum fields}

With all the tools in hand, we can now investigate the relative observables for the Poincar{\'e} group 
\begin{equation}
G = T(1,3) \rtimes SO(1,3)
\end{equation}
in some detail. Writing $\vec{0}$ for the trivial translation and $e \in SO(1,3)$ for the trivial Lorentz transformation we notice that a generic group element $(\vec{x},\Lambda) \in G$ can be written
\begin{equation}
(\vec{x},\Lambda)=(\vec{x},e).(\vec{0},\Lambda),
\end{equation}
which allows us to write relative observables as\footnote{See \cite{mazzucchi_observables_2001} and \cite{ali_systems_1998} for the study of covariant POVMs on the Poincar{\'e} group.}
\begin{align} \begin{split}
    \yen^\R(A) &= \int_{G} A.(x,\Lambda) \otimes d\E_\R(x,\Lambda)\\
    &= \int_G A(x).\Lambda \otimes d\E_\R(x,\Lambda),
\end{split} \end{align}
where $A(x) := A.(x,e)$ and $A.\Lambda := A.(\vec{0},\Lambda)$. As already noted, the $\Y^\R$ integration is insensitive to the choice of the identity in the torsor of frames. At the same token, we can now identify the translation group with the Minkowski spacetime, written $\mathbb{M} \cong T(1,3)$, and interpret
\begin{equation}
\hat{A}: \mathbb{M} \ni x \mapsto A(x) \in B(\his)
\end{equation}
as a \emph{quantum field}. To support this interpretation, we consider the \emph{reduced} setup with a frame defined on $T(1,3) \subset G$, and prepared in a state $\omega \in \St(\hir)$ with the corresponding probability distribution supported in the region $U \subseteq\mathbb{M} \cong T(1,3)$. The notion of a local smeared field observable is then recovered via
\begin{equation}\label{locaqft}
    \Y^\R_\omega (A) = \int_U \hat{A}(x) \h d\mu^{\mathsf{F}_\R}_\omega(x).
\end{equation}
Thus the local QFT observables arise in the context of a quantum reference frame on a reduced symmetry group. We call them \emph{relational local observables}. \begin{comment}
Treating the action of the Poincar{\'e} group $G$ on $\mathbb{M}$ as a family of diffeomorphisms and performing a change of variables the \emph{covariance} of relational local observables is established in the following sense
\begin{align} \begin{split}
    \Y_\omega^{\R}(A).(\vec{v},\Lambda) &= \int_U \hat{A}(x).(\vec{v},\Lambda) \h d\mu_\omega^{\mathsf{F}_\R}(x)\\
    &= \int_U \hat{A}(\Lambda x+\vec{v}) \h d\mu_\omega^{\mathsf{F}_\R}(x)\\
    &= \int_{U.(\vec{v},\Lambda)} \hat{A}(x) \h d\mu_\omega^{\mathsf{F}_\R}(\Lambda^{-1}(x-\vec{v})),
\end{split} \end{align}
where $(\vec{v},\Lambda) \in G$ is arbitrary and we have used transitivity of the action of $G$ on $\mathbb{M}$.
\end{comment}
To what extent the relational local observables are compatible with the Wightman axioms and the axioms of algebraic QFT, or relational versions thereof, is being investigated in a separate work. The concept is generalized to the context of general principal bundles~in~\ref{genrellocobs}.

Notice also, that if we insist on giving physical relevance only to quantities arising through our relational procedure, the region $U \subseteq\mathbb{M}$ becomes an entity only \emph{relationally}, when \eqref{locaqft} is non-zero, which can be interpreted as a possibility for an `interaction', happening in $U$, between the system and the frame. In the sequel, when we deal with frames in the context of principal bundles, they will be defined together with subsets of the space-time manifold on which such distributions can be supported, making the frame-relationality of relational local observables even more apparent. We also note here that, like in ordinary QFT, altering the field at a single point (or any other measure-zero subset of $\mathbb{M}$) will leave the local observables intact. Thus, the fields entering the relativization procedure can, and perhaps should be considered as \emph{equivalence classes} of such operator-valued functions. Further details on the operators arising as integrals with respect to operator-valued measures are given in the next section.

Finally, let us note that there seem to be no obstacles to introducing additional gauge symmetry group $H$ so that the relative operators take the form
\begin{equation}
    \Y^\R(A) = \int_{\mathbb{M} \times SO(1,3) \times H} \hat{A}(x).(\Lambda,h) \otimes d\E_\R(x,\Lambda,h).
\end{equation}
This way a `probabilistic local gauge fixing' can be attained upon restriction on a state preparation of the frame in the sense that, in the case of a frame based on $T(1,3) \times H$, we have
\begin{equation}
    \Y^\R_\omega (A) = \int_{\mathbb{M} \times H} \hat{A}(x).h \h d\mu^{\E_\R}_\omega (x,h).
\end{equation}
We refer to the formalism steaming from these considerations, extended beyond the special-relativistic realm in the sequel, as the \emph{relational} approach to Quantum Field Theory (RQFT).

\section{Mathematical Intermezzo}

To be able to apply the formalism presented here in the context of curved spacetimes and beyond, we now need to make a small detour into the relevant mathematics. Less mathematically inclined readers are welcome to just take a look at the formulas \eqref{eq:intopchvar}, \eqref{eq:intopeqch} and the Thm. \ref{thm:intop} below that establish the tools needed in the sequel.

\subsection{Operator-valued integration}\label{ovi}
In this subsection, we introduce operator-valued integration, which is an entirely new tool that allows us to push the formalism of operational QRFs to the realm of gauge theories in the next section.

The integral with respect to a POVM defining the relativization map \eqref{eq:yen}, originally developed in the context of locally compact second countable Hausdorff groups and then generalized to finite \cite{glowacki_quantum_2024} and compactly stabilized \cite{fewster_quantum_2024} homogeneous spaces, is a special case of a very general instance of an operator-valued integral that we now introduce. Given a bounded and continuous (in ultraweak topology)\footnote{The ultraweak topology is the coarsest one making all the maps $A \mapsto \tr[\omega A]$ continuous. We also note here that such continuity and boundedness is not a necessary condition for an operator-valued integral (see below) to be defined; the functions $f$ such that all integrals \eqref{integrals} converge are called $\E_\R$-integrable in \cite{glowacki_operator-valued_2024}, where the general theory is being developed.} operator-valued function $f: \Sigma \to B(\his)$, where $\Sigma$ is a topological measurable space, any POVM $\E_\R:{\rm Bor}(\Sigma) \to \Eff(\hir)$ defines a unique operator in $B(\hisr)$ by specifying its expectation values on the product states to be
\begin{equation}\label{eq:integrals}
\tr[\rho \otimes \omega \int_\Sigma f \otimes d\E_\R] := \int_\Sigma \tr[\rho f(x)] d\mu^{\E_\R}_\omega.
\end{equation}

At the heart of the definition of such operators, which we refer to as \emph{operator-valued integrals}, lies the Banach duality $B(\hi) \cong \T(\hi)^*$ which, together with the Hahn-Banach Theorem, assures that a bounded operator on $\hi$ is uniquely specified by its expectation values on any linearly dense subset of $\T(\hi)$, which here is taken to be the product states on $\hisr$. Since the functions
\begin{equation}
f_\rho: \Sigma \ni x \mapsto \tr[\rho f(x)] \in \Cn
\end{equation}
are continuous and bounded and the measures $\mu^{\E_\R}_\omega$ are finite, the integrals \eqref{eq:integrals} will all converge. Moreover, since the definition of operator-valued integrals is based on Lebesgue integration theory, one easily shows that given a continuous map $\varphi: \Sigma \to \Sigma'$ we can perform a change of variables
    \begin{equation}\label{eq:intopchvar}
        \int_\Sigma f \circ \varphi \otimes d\E_\R = \int_{\varphi(\Sigma)} f \otimes d(\E_\R \circ \varphi^{-1}).
    \end{equation}
Moreover, for an arbitrary channel $\psi: B(\hir) \to B(\hirprim)$, we have
\begin{equation}\label{eq:intopeqch}
    \int_\Sigma f \otimes d(\psi \circ \E_\R) = \id_{\his} \otimes \psi \left(\int_\Sigma f \otimes d\E_\R\right).
\end{equation}

Since this is the first time these claims appear in the literature, we now phrase them as a theorem.

\begin{theorem}\label{thm:intop}
    For any ultraweakly continuous function $f: \Sigma \to B(\his)$ and POVM $\E_\R: {\rm Bor}(\Sigma) \to~B(\hir)$ there exists a unique bounded operator on $\hisr$, written 
    \[
    \int_\Sigma f(x) \otimes d\E_\R(x),
    \]
    such that for all $\rho \in \S(\his)$ and $\omega \in \S(\hir)$ we have
    \begin{equation}\label{integrals}
        \tr[\rho \otimes \omega \h \int_\Sigma f(x) \otimes d\E_\R(x)] = \int_\Sigma f_\rho \h d\mu^{\E_\R}_\omega,
    \end{equation}
    and the properties \eqref{eq:intopchvar} and \eqref{eq:intopeqch} follow.
\end{theorem}

\begin{proof}
    See App. \ref{app:proof}.
\end{proof}

\subsection{Operator-valued functions and PDEs}\label{sec:PDEs}

Before we go back to RQFT in the next section, it may be insightful to introduce one more concept -- that of a differential equation for operator-valued functions. We stress here that this section, being widely independent from the rest of the paper, is more speculative in the sense that the presented ideas have not yet been investigated enough to assure their validity or relevance. Nevertheless, they seem to provide a clear connection between gauge symmetries and differential equations in the context of RQFT, which we find illuminating and potentially useful. They share the Banach duality for trace-class/bounded operators as a common core with the approach to operator-valued integration briefly described in Sec. \ref{ovi}. We see this approach as a potential alternative to the standard distributional treatment (see e.g. \cite{rejzner_algebraic_2016}).

We will be interested in partial differential equations (PDEs) that can be formalized as operators
\begin{equation}
    T: C^\infty(U) \supset V \to C^\infty(U),
\end{equation}
where $U \subseteq \mathcal{M}$ is a (bounded) domain in a manifold $\mathcal{M}$ and $V \subseteq C^\infty(U)$ a subspace on which $T$ is defined. Now consider an operator-valued function
\begin{equation}
    \hat{\phi}: U \to B(\his)
\end{equation}
with the property that for any state $\rho \in \St(\his)$, the map
\begin{equation}
    \hat{\phi}_\rho: U \ni p \mapsto \tr[\rho \h \hat{\phi}(p)] \in \Cn
\end{equation}
is in $V$; denote the space of such functions by $\hat{V}$. The differential operator $T$ can then be `lifted' to 
\begin{equation}
    \hat{T}: C^\infty(U,B(\his)) \supset \hat{V} \to C^\infty(U,B(\his)),
\end{equation}
where $C^\infty(U,B(\his))$ denotes the space of operator-valued functions such that all $\hat{\phi}_\rho$ are smooth. Indeed, we can take $\hat{T}(\hat{\phi})(p) \in B(\his)$ to be defined by\footnote{Notice that $T(\hat{\phi}_\rho)$ being smooth needs to be bounded on a bounded domain, making RHS of \eqref{ovpde} finite. Just like in defining the integrals with respect to operator-valued measure, we exploit here the Banach duality $B(\hi) \cong \T(\hi)^*$, which together with the Hahn-Banach Theorem assures that a bounded operator is uniquely specified by its expectation values on states.}
\begin{equation}\label{ovpde}
    \tr[\rho \h \hat{T}(\hat{\phi})(p)] :=  T(\hat{\phi}_\rho)(p).
\end{equation}
We are using the functionals $A \mapsto \tr[\rho \h A]$ in a way analogous to how atlases are used in Differential Geometry -- they allow us to lift the concepts from the theory of PDEs for scalar-valued functions to those with values in operator algebras.

Now assume there is an action of a (Lie) group $H$ on the space of solutions ${\rm ker}(T) \subset V$. We will then have a corresponding action on ${\rm ker}(\hat{T}) \subset \hat{V}$ with $\hat{\phi}.h$ defined via
\begin{equation}
    \tr[\rho \h (\hat{\phi}.h)(p)] := (\hat{\phi}_\rho.h)(p),
\end{equation}
for any $\rho \in \St(\his)$, where on the RHS we invoked the action of $H$ on ${\rm ker}(T)$. The group of symmetries of the space of solutions of a given PDE then translates to an action of the same group on the operator algebra in which the solutions to the `lifted' PDE take values. 

\section{Quantum Frames and Fields}\label{sec:QFaF}

In this section, we generalize the operational approach to QRF to the realm of gauge theories.

\subsection{QRFs on principal bundles}\label{qrfPB}

To generalize the ideas presented in Sec. \ref{QRFs}, and also get a better conceptual understanding of the emerging relational formalism for relativistic quantum fields, we need to generalize the relativization procedure. Thanks to the results of operator-valued integration described in Sec. \ref{ovi}, we can consider frame observables on arbitrary topological spaces; the setup useful for our present discussion is that of a \emph{principal} $H$-\emph{bundle}, with $H$ a (topological) group. Such a bundle is given by a (surjective) map $\pi: B \to \mathcal{M}$, where $B$ and $\mathcal{M}$ are topological spaces (the latter to be thought of as spacetime), $H$ acts on $B$ (on the right), and for any $p \in \mathcal{M}$ there is a neighbourhood $p \in U \subset \mathcal{M}$ such that
\begin{equation}\label{loctriv}
    \pi^{-1}(U) \cong U \times H,
\end{equation}
with $b.g = (p,hg)$ for all $b = (p,h) \in \pi^{-1}(U)$ and $g \in H$. A homeomorphism \eqref{loctriv} can be specified by means of a \emph{local section}, i.e, a (continuous) map
\begin{equation}
    \sigma: U \to B,
\end{equation}
such that $\pi \circ \sigma (p) = p$ for all $p \in U$. This is achieved by taking $b =(\pi(b),h_\sigma(b))$, where $h_\sigma(b) \in H$ is the unique group element such that $b.h_\sigma(b) = \sigma(\pi(b))$. In (other) words, the group element associated with $b$ describes the orientation of $b$ with respect to $\sigma$.

A suitable definition for a QRF in the context of principal bundles seems to be the following:

\begin{definition}
    A \emph{quantum reference frame on a principal $H$-bundle} $\pi: B \to \mathcal{M}$ is an $H$-covariant POVM
\begin{equation}
    \E_\R: {\rm Bor}(\pi^{-1}(U)) \to \Eff(\hir),
\end{equation}
together with the choice of a local section $\sigma: U \to B$.
\end{definition}

The POVM $\E_\R$ of the frame will be referred to as the \emph{frame observable}; a frame will be called \emph{sharp} if all the effects $\E_\R(X)$ are projections, and \emph{ideal} if the Hilbert space of the frame is $\hir = L^2(\pi^{-1}(U))$ with the natural action of $H$ and the covariant POVM assigns to the Borel subsets the operators of multiplication by their characteristic functions
\[
    P: {\rm Bor}(\pi^{-1}(U)) \ni Y \mapsto M_{\chi_Y} \in \Eff(L^2(\pi^{-1}(U))).
\]

Let us point here to some subtleties regarding the subset $U \subset \mathcal{M}$. Firstly, in contrast to its appearance in the context of QRFs for Poincar{\'e} group in \eqref{locaqft}, it is now part of the \emph{definition} of the frame, and not a property of the state $\omega \in \mathcal{D}(\hir)$ as before; we see the point of view arising here from the need of choosing local trivialization of the bundle -- which is already granted globally for the trivial Poincar{\'e} bundle -- to be better aligned with the operational principles. We believe the set $U$ should be thought of as the sample space corresponding to the measurements of time differences and spacial distances available in the concrete experimental scenario, and as such restricted to the laboratory where the observations take place. Indeed, the very manifold $\mathcal{M}$ should perhaps be thought of as only arising in the operational description when it can be covered by the sample spaces of the frames considered. From this point of view, the sets $U$ should perhaps be considered bounded, which is aligned both with relating the relational local observables to their established distributional definitions (see \eqref{smearedqf2} below), and the local treatment of PDEs for operator-valued functions as described in Sec. \ref{sec:PDEs}.

%We also note here that, to the best of our knowledge, the existence of covariant POVMs on general (not trivial) principal bundles has not yet been investigated in the literature, which places the presented approach to relational QFT in the speculative domain. However, we can not see obstacles for such objects and are planning to attempt to construct them in future work. 

\subsection{Relativization for quantum fields}

Frames on principal bundles are well-suited to relativize quantum fields in the following sense:

\begin{definition}
    Given a frame $\R = (\E_\R,\sigma)$ on a principal bundle $\pi: B \to \mathcal{M}$, the \emph{relativization map} $\Y^\R$ takes operator-valued functions $\hat{\phi}: \mathcal{M} \to B(\his)$ to $H$-invariant operators in $B(\hisr)^H$~via
\begin{equation}\label{relqf}
    \Y^\R: \hat{\phi} \mapsto \int_{\pi^{-1}(U)} \hat{\phi}(\pi(b)).h_\sigma(b) \otimes d\E_\R(b),
\end{equation}
where $h_\sigma: \pi^{-1}(U) \to H$ is fixed by $b.h_\sigma(b) = \sigma(\pi(b))$.
\end{definition}

The function $b \mapsto \hat{\phi}(\pi(b)).h_\sigma(b)$ is easily assured to be (ultraweakly) continuous and bounded if only the field $\hat{\phi}$ and the action of $H$ on $B(\his)$ are such, and thus $\Y^\R(\hat{\phi})$ can be defined by \eqref{eq:integrals}. The image of $\Y^\R$ lies in the $H$-invariant subalgebra as easily confirmed by the following calculation
\begin{align} \begin{split}
    &\Y^\R(\hat{\phi}).g = \int_{\pi^{-1}(U)} \hat{\phi}(\pi(b)).(h_\sigma(b)g) \otimes d\E_\R(bg)=\\
    &\int_{\pi^{-1}(U)} \hat{\phi}(\pi(bg)).(h_\sigma(bg)) \otimes d\E_\R(bg) = \Y^\R(\hat{\phi}),
\end{split} \end{align}
where $g \in H$ is arbitrary and we have used $H$-covariance of $\E_\R$ and the fact that $H$ acts on $B$ in a fiber-wise fashion. The domain of $\Y^\R$ is thought of consisting of fields that satisfy some differential equation (with $H$ as the group of symmetries of solutions) in the sense of \eqref{ovpde}. 

Upon relativization, the operators $\hat{\phi}(p)$ are then rotated accordingly to the orientation of the frame, understood relative to the section $\sigma: U \to B$, and integrated over $\pi^{-1}(U)$ with $\E_\R$, resulting in an operator on the composite system just like in the case of QRFs on groups. The \emph{global} gauge invariance in the sense of $H$-invariance is assured, while \emph{local} gauge is fixed by the section $\sigma$, which allows for relational and local `coupling' of the frame to the quantum field.

%We also note here that we always have an action of the group of homeomorphisms, or diffeomorphisms if manifold structure is assumed on $\mathcal{M}$, on the space of operator-valued functions given by pre-composition. The relativization is constant on the orbits of this action as follows from a simple change of variables. Indeed,  for any homeomorphism $\varphi: U \to U$ we have\footnote{Strictly speaking, to perform the change of variables as provided by Thm. \ref{thm:intop} we need to lift $\varphi: U \to U$ to the domain of integration, which can be written explicitly as $\tilde{\varphi}: \pi^{-1}(U) \ni (\pi(b),h_\sigma(b)) \mapsto (\varphi(\pi(b)),h_\sigma(b)) \in \pi^{-1}(U)$.}
%\begin{align} \begin{split}
    %\Y^\R(\hat{\phi} \circ \varphi) =&\\
    %\int_{\pi^{-1}(U)} \hat{\phi} \circ \varphi(\pi(b)).h_\sigma(b) \otimes d\E_\R(\pi(b),h_\sigma(b)) =&\\
    %\int_{\pi^{-1}(\varphi(U))} \hat{\phi}(\pi(b)).h_\sigma(b) \otimes d\E_\R(\varphi(\pi(b)),h_\sigma(b)) =&\\
    %\Y^\R(\hat{\phi}).&
%\end{split} \end{align}
%In the case when $\mathcal{M}$ is a spacetime manifold, this can be thought of as the form of local \emph{diffeomorphism-invariance} that the formalism obeys.

\subsection{Restriction for quantum fields}\label{subsec:resqf}

We now consider restricting the relative descriptions of quantum fields upon a choice of state preparation of the frame. In general this gives
\begin{equation}
    \Y^\R_\omega(\hat{\phi}) = \int_{\pi^{-1}(U)} \hat{\phi}(\pi(b)).h_\sigma(b) \h d\mu^{\E_\R}_{\omega} \in B(\his).
\end{equation}

The analogue of localization at $e \in G$ for a QRF on a group seems to be localization at a point in $\sigma(\mathcal{M})$. To be precise, we consider a QRF $\R = (\E_\R,\sigma)$ and a sequence of states $\omega_n(p)\in \St(\hir)$ such that $\lim_{n \to \infty} \mu^{\E_\R}_{\omega_n(p)} (b) = \delta_{\sigma(p)}$. In such a case we get
\begin{equation}
\lim_{n \to \infty}\Y^\R_{\omega_n(p)}(\hat{\phi}) = \hat{\phi}(p).
\end{equation}

We note that this limit can \emph{not} be taken if $\lim_{n\to \infty}\omega_n$ is to remain a normal quantum state \cite{carette_operational_2023}; the field operators at a single point remain undetermined.

\subsection{Reduction for quantum fields}\label{subsec:redqf}

The notion of a reduction of a principal bundle suitable for our purposes is the following.

\begin{definition}
    An $H_\R$-principal bundle $\pi_\R: B_\R \to \mathcal{M}_\R$ is a \emph{sub-bundle} of the $H$-principal bundle $\pi: B \to \mathcal{M}$ if $H_\R < H$ and we have a pair of embeddings such that
\[
\begin{tikzcd}
    B_\R \arrow[d, "\pi_\R"'] \arrow[r, "i", hook] & B \arrow[d, "\pi"] \\
    \mathcal{M}_\R \arrow[r, "j"] & \mathcal{M}
\end{tikzcd}
\]
commutes.
\end{definition}
It follows that $i$ above is $H_\R$-equivariant. Given a QRF
\begin{align} \begin{split}
&\mathsf{F}_\R: {\rm Bor}(\pi_\R^{-1}(U_\R)) \to \Eff(\hir),\\
&\gamma_R: \mathcal{M}_\R \supset U_\R \to B_\R,
\end{split} \end{align}
we can extend it to a frame on $\pi: B \to \mathcal{M}$ by taking
\begin{align} \begin{split}
\E_\R &:= \mathsf{F}_\R \circ i^{-1}: {\rm Bor}(\pi^{-1}(j(U_\R))) \to \Eff(\hir),\\
\sigma &:= i \circ \gamma_R \circ j^{-1}: j(U_R) \to B;
\end{split} \end{align}
%then $\E_\R(X) = \E_\R(X \cap i(B_\R))$ and 
The fields on $U_\R$ can be then relativized via
\begin{align} \begin{split}
&\Y^\R(\hat{\phi}|_{j(U_\R)})=
\\&= \int_{i(\pi_\R^{-1}(U_\R))} \hat{\phi}(\pi(b)).h_{i \circ \gamma_\R \circ j^{-1}}(b) \otimes d(\mathsf{F}_\R \circ i^{-1})(b)\\
    &= \int_{\pi^{-1}_\R(U_\R)} \hat{\phi}(\pi(i(b))).h_{\gamma_\R}(b) \otimes d\mathsf{F}_\R(b)\\
    &= \int_{\pi^{-1}_\R(U_\R)} \hat{\phi} \circ j (\pi_\R(b)).h_{\gamma_\R}(b) \otimes d\mathsf{F}_\R(b),
\end{split} \end{align}
where we have performed a change of variables associated to $i$. Such operators are invariant under global $H_\R$-gauge transformations. We understand reduction as a way to introduce a frame that admits a smaller gauge symmetry group and/or is defined over a smaller base space to be used to relativize fields defined on $\mathcal{M}$ and subject to an action of the gauge group $H$. 

\subsection{Relational local observables}\label{genrellocobs}

As an example of a frame based on a reduced bundle, consider an $\{e\}$-principal bundle
\begin{equation}
\pi_\R = {\rm Id}_U: B_\R \cong U \to U \subseteq \mathcal{M}
\end{equation}
with the embeddings $i = \sigma|_U: U \to B$ and $j: U \hookrightarrow \mathcal{M}$. Any~POVM
\begin{equation}
{\mathsf{F}_\R}: {\rm Bor}(U) \to B(\hir)
\end{equation}
will then specify a QRF, since it will be trivially covariant. The relative operators taken with respect to such a frame take the form
\begin{align} \begin{split}
    \Y^\R(\hat{\phi})
    = \int_U \hat{\phi}(p) \otimes d\mathsf{F}_\R(p),
\end{split} \end{align}
and the smeared field observables are again recovered under the restriction
\begin{equation}\label{smearedqf2}
    \Y^\R_\omega(\hat{\phi}) = \int_U \hat{\phi}(p) \h d\mu^{\mathsf{F}_\R}_\omega(p) \in B(\his).
\end{equation}
We refer to such operators as \emph{relational local observables}, while the algebras they generate via (ultraweak closure)
\begin{equation}
    \mathcal{A}^\R := \{\Y^\R_\omega(\hat{\phi}) \h | \h \omega \in \St(\hir)\}^{cl}
\end{equation}
are called \emph{relational local algebras}\footnote{Notice that $\mathcal{A}^\R$ is implicitly associated with $U \subseteq \mathcal{M}$ on which $\R$ is based.}; their properties under frame changes are being investigated in separate work. More examples of relational local observables will be introduced in the next section in the context of relativistic fields and frames on principal Lorentz bundles.

\subsection{Changing frames for quantum fields}

Here we generalize the concept of an \emph{external} frame transformation to the context of QRFs on principal bundles. Extending the notion introduced in \cite{glowacki_relativization_2024} we propose the following definition.

\begin{definition}\label{def:frmmor}
Consider a \emph{pair of frames} on a principal $H$-bundle $\pi: B \to \mathcal{M}$,~i.e, 
    \begin{itemize}
        \item a pair of $H$-covariant POVMs
        \[\begin{split}
            \E_\R&: {\rm Bor}(\pi^{-1}(U)) \to \Eff(\hir),\\
            \E_{\R'}&: {\rm Bor}(\pi^{-1}(U')) \to \Eff(\hirprim),
        \end{split}\]
        where $U,U' \subseteq \mathcal{M}$, and
        \item a pair of sections $\sigma: U \to B$, $\sigma': U' \to B$.
    \end{itemize}
Then a pair $(\psi,\theta)$, where
\begin{itemize} 
    \item $\psi$ is a channel between the frame algebras, i.e,
    \[
        \psi: B(\hir) \to B(\hirprim), \text{ and }
    \]
    \item $\theta$ is a fiber-preserving (continuous) map
    \[
        \theta: \pi^{-1}(U) \to \pi^{-1}(U')
    \]
\end{itemize}
constitutes a \emph{frame morphism}, written
\[
(\psi,\theta): (\R,\sigma) \to (\R',\sigma'),
\]
if the following compatibility conditions hold:
\begin{itemize}
    \item[1)] %$\theta$ is compatible with the choices of the sections, i.e, 
    the following diagram commutes
    \[
\begin{tikzcd}
    \pi^{-1}(U) \arrow[r, "\theta "] & \pi^{-1}(U') \\
    U \arrow[u, "\sigma"] \arrow[r, "\varphi_\theta"] & U' \arrow[u, "\sigma'"']
\end{tikzcd},
\]
where $\varphi_\theta:= \pi \circ \theta \circ \pi^{-1}$, and
\item[2)] the frame observables are related via
\[
    \E_{\R'} = \psi \circ \E_\R \circ \theta^{-1}.
\]
\end{itemize}
\end{definition}

Notice that for $(\psi,\theta)$ to be a frame morphism, $\psi$ needs to be equivariant and $\theta$ surjective, and that $\varphi_\theta$ above is well-defined since $\theta$ preserves fibers.

The definition above combines different ways in which the reference frames can be related. For example, taking $\varphi_\theta = {\rm Id}_U$, a morphism $({\rm Id}_\R,\theta)$ can be understood as relating the different gauge choices associated with $\sigma$ and $\sigma'$, while keeping the rest of the setup intact. As we will see in the sequel, this is a \emph{strict generalization} of the concept of a local coordinate transformation. More generally, if we allow $U$ and $U'$ to be distinct, $\theta$ provides means for understanding the second frame to be the first one (if we keep $\psi = {\rm Id}_\R$) but `relocated' to $U'$. On the other hand, if the quantum systems on which the frames are based are distinct but we keep $U=U'$, the morphism $(\psi,{\rm Id}_{\pi^{-1}(U)})$ describes a change of the physical system used as a reference, without altering the choice of the local gauge. As a last comment in this context, consider $U$ and $U'$ overlapping, i,e, $U \cap U' \neq \emptyset$. Then $\theta$ provides a gluing for the sections $\sigma$ and $\sigma'$ on $U \cap U'$.

Frame morphisms allow for external frame transformations mapping between the relative observables defined with respect to different frames. Indeed, given a frame morphism $(\psi,\theta): (\R,\sigma) \to (\R',\sigma')$ the relative operator $\Y^{\R'}(\hat{\phi})$ can be constructed from $\Y^\R(\hat{\phi})$ via
\begin{align} \begin{split}
    \Y^{\R'}(\hat{\phi})
    &= \int_{\pi^{-1}(U')} \hat{\phi}(\pi(b)).h_{\sigma'}(b) \otimes d\E_{\R'}(b)\\
    &= \int_{\theta(\pi^{-1}(U))} \hat{\phi}(\pi(b)).h_{\sigma'}(b) \otimes d(\psi \circ \E_{\R} \circ \theta^{-1})(b)\\
    &= \int_{\pi^{-1}(U)} \hat{\phi}(\pi(\theta(b))).h_{\sigma' \circ \varphi}(\theta(b)) \otimes d(\psi \circ \E_{\R})(b)\\
    &= \int_{\pi^{-1}(U)} \hat{\phi}(\varphi_\theta (\pi(b))).h_\sigma(b) \otimes d(\psi \circ \E_{\R})(b)\\
    &= (\id_\S \otimes \psi) \int_{\pi^{-1}(U)} \hat{\phi} \circ \varphi_\theta (\pi(b)).h_{\sigma}(b) \otimes d\E_{\R}(b)\\
    &=(\id_\S \otimes \psi)\left(\Y^{\R}(\hat{\phi} \circ \varphi_\theta)\right),
\end{split} \end{align}
where we have first used surjectivity of $\theta$ and the relation between the frame observables, then performed the change of variables dictated by $\theta$, next used its properties as a bundle map compatible with the sections to rephrase the expression under the integral, and finally took the channel relating the frames out of the integral.

Accordingly to the discussion below the Def. \ref{def:frmmor}, an external QRF transformation can capture a gauge transformations, a frame relocation, a change of a system used as a reference, and any combination thereof. We believe that covariance with respect to quantum reference frames may be best understood in the context of external quantum reference frames transformations as defined above.

\section{QRFs and Curved Spacetimes}\label{RQFTcurved}

In this section, we apply the gauge-theoretic setup for QRFs introduced above to the case of quantum fields with local Lorentz gauge symmetry. They can be described relative to quantum frames associated with the symmetry structure of a fixed curved spacetime manifold.

\subsection{Lorentzian geometry and gauge}

To describe quantum fields relative to quantum frames on a curved background, we need to relax the Poincar{\'e} symmetry assumed to underlie the description, which is only appropriate in the absence of gravity. To this end, we note the following. One way to understand the role of the gravitational field in GR is to say that it specifies the class of inertial frames at each point on $\mathcal{M}$ (which is now taken to be a $4$-dimensional, time-oriented manifold). To be precise, specifying a metric tensor on $\mathcal{M}$ is \emph{equivalent} to declaring, at each $p \in \mathcal{M}$, which bases of the tangent space $T_p\mathcal{M}$ will be \emph{pseudo-orthonormal}, i.e, in which the metric tensor $g_{\mu\nu}(p)$ will take the flat form of $\eta={\rm diag}(-1,1,1,1)$. Clearly, for any two bases in $T_p\mathcal{M}$ related by a Lorentz transformation\footnote{We restrict to \emph{special} orthogonal transformations to assure all the frames on the orbit to preserve time-orientation of $\mathcal{M}$.}, they will either both be \emph{pseudo-orthonormal}, or neither of them and if they are, such Lorentz transformation between them is unique. This brings us to considering a gauge structure associated to a space-time $(\mathcal{M},g_{\mu\nu})$ to be a \emph{principal} $SO(1,3)$-\emph{bundle}, written $\pi_{g_{\mu\nu}}: B_{g_{\mu\nu}} \to \mathcal{M}$. It will now replace the Poincar{\'e} group, which can be seen as a \emph{trivial} $SO(1,3)$-bundle $T(1,3) \rtimes SO(1,3) \to \mathbb{M}$, in our considerations.

\subsection{Frames on curved spacetimes}

Recall that a QRF on a principal $H$-bundle $\pi: B \to \mathcal{M}$ is an $H$-covariant POVM on the subset of the total space
\begin{equation}
    \E_\R: {\rm Bor}(\pi^{-1}(U)) \to \Eff(\hir),
\end{equation}
together with a section $\sigma: U \to B$. In the case of an $SO(1,3)$-bundle of $(\mathcal{M},g_{\mu\nu})$, such a section is often referred to as a local \emph{tetrad field}; it corresponds to a choice of local coordinates on $U$ in which $g_{\mu\nu}$ is diagonal. Thus, a QRF \emph{on a spacetime} $(\mathcal{M},g_{\mu\nu})$ is a covariant POVM on the (subset of the) $SO(1,3)$-bundle corresponding to $(\mathcal{M},g_{\mu\nu})$ together with a local tetrad field, i.e, it is given by
\begin{equation}
    \E_\R: {\rm Bor}(\pi_{g_{\mu\nu}}^{-1}(U)) \to \Eff(\hir), \haa \sigma_T: U \to B_{g_{\mu\nu}}.
\end{equation}

We note here that, just like in the case of fields on a Minkowski spacetime, additional gauge degrees of freedom are easily introduced by considering a principal bundle with a bigger symmetry group.

\subsection{Reductions of relativistic frames}

Applying the ideas described in Subsec. \ref{subsec:redqf}, we can take $\E_\R = {\mathsf{F}_\R} \circ \sigma_T^{-1}$ with ${\mathsf{F}_\R}$ based on $\mathcal{M}$ and get smeared quantum fields as a result of restriction as in \eqref{smearedqf2}. We will now take this idea further and consider a \emph{path} $\gamma: \Reals \to \mathcal{M}$, together with a POVM $\mathsf{F}_\R$ defined on $\Reals$. The frame observable can now be taken to be
\begin{equation}
    \E_\R = \mathsf{F}_\R \circ (\sigma_T \circ \gamma)^{-1},
\end{equation}
with the property that $\E_\R(X) = \E_\R(X \cap \sigma_T \circ \gamma (\Reals))$, so that the tetrad field along $\gamma$ is always aligned with $\sigma$. Upon restriction we then get
\begin{equation}
\Y^\R_\omega(\hat{\phi}|_{\gamma(\Reals)}) = \int_\Reals \hat{\phi}(\gamma(t)) \h d\mu^{\mathsf{F}_\R}_\omega(t),
\end{equation}
which can be understood as an average value of the field observable over the path, weighted accordingly to the probability of the frame in use being localized in the specific points along it. Using the notation of Subsec. \ref{subsec:redqf}, this setup corresponds to $\pi_\R ={\rm Id}_\mathbb{R}$, $i=\sigma_T \circ \gamma$ and $j=\gamma$.

We can also consider a frame localized \emph{above} a path, i.e., allow for non-trivial orientation of the tetrad along $\gamma$. To this end, we take $i = \bar{\gamma}: \Reals \to B_{g_{\mu\nu}}$ such that $\pi \circ \bar{\gamma} = \gamma$,
\begin{equation}
\mathsf{F}_{\bar{\R}}: {\rm Bor}(\Reals) \to \Eff(\hir)
\text{, and }
\E_\R = \mathsf{F}_{\bar{\R}} \circ \bar{\gamma}^{-1}.
\end{equation}
Upon restriction, writing $\Lambda_{\sigma_T}(t) := \Lambda_{\sigma_T}(\bar{\gamma}(t))$ this gives
\begin{equation}
\Y^\R_\omega(\hat{\phi}|_{\gamma(\Reals)}) = \int_\Reals \hat{\phi}(\gamma(t)).\Lambda_{\sigma_T}(t) \h d\mu^{\mathsf{F}_{\bar{\R}}}_\omega(t).
\end{equation}

To make the orientation of the tetrad field along the path indefinite, we consider a frame observable defined on a sub-bundle given by
\begin{equation}
\pi_\R: B_{g_{\mu\nu}}|_{\gamma(\Reals)} \cong \Reals \times SO(1,3) \to \gamma(\Reals) \cong \Reals.
\end{equation}
This way the restricted observables will take the form
\begin{equation}
\Y^\R_\omega(\hat{\phi}|_{\gamma(\Reals)}) = \int_{\Reals \times SO(1,3)} \hat{\phi}(\gamma(t)).\Lambda \h d\mu^{\E_\R}_\omega(t,\Lambda).
\end{equation}
Reducing further to the sub-bundle for rotations $SO(3) \subset SO(1,3)$ we can describe a QRF that is `stationary' along a given path. This generalizes the setup employed in \cite{fewster_quantum_2024}, where a QRF along a path in de Sitter spacetime is considered.\footnote{We suspect that the presented formalism is capable of modelling relativistic detectors for quantum fields in very general measurement scenarios, filling the gap between abstract algebraic treatment \cite{fewster_measurement_2023} and various models such as that of the Unruh-DeWitt detector, providing the possibility of indefinite localization.}

\subsection{Isometric frame transformations}

Recall that the data needed for an external frame transformation consists of an equivariant channel and a fiber preserving map allowing to expression of the frame observable $\E_{\R'}$ with $\E_\R$ in that
\begin{equation}
    \E_{\R'} = \psi \circ \E_\R \circ \theta,
\end{equation}
and compatible with the sections associated with the frames in the sense that
\begin{equation}
    \theta \circ \sigma = \sigma' \circ \varphi_\theta.
\end{equation}
This is the case e.g. when $\varphi_\theta$ is an isometry and $\theta$ the transformation of the bundle associated with the differential $T\varphi_\theta: TU \to TU'$. An external frame change for relativistic quantum fields can then be achieved via an equivariant channel and a \emph{space-time isometry} (lifted to the level of the Lorentz bundle).

\section{Non-inertial QRFs}\label{noniner}

In this section, we explore relational quantum fields on spacetimes without a predetermined metric structure, which allows for more general frames to be considered. They admit a natural interpretation as being non-necessary \emph{inertial} with respect to the Lorentzian geometry specified by their sections.

\subsection{GR with frame bundles}

Consider a ($4$-dimensional) manifold $\mathcal{M}$ \emph{without} any particular metric tensor being specified. The \emph{frame bundle} $\pi: L\mathcal{M} \to \mathcal{M}$ is a principal $GL(4,\Reals)$-bundle where elements of a fiber $L_p\mathcal{M}$ represent choices of a basis for the tangent space $T_p\mathcal{M}$, and thus are all related by invertible linear transformations, in a unique way. Any local coordinate chart
\begin{equation}
{\rm x}: \mathcal{M} \supset U \ni p \mapsto \{x^\mu(p)\} \in V \subset \Reals^4
\end{equation}
provides a local section of $L\mathcal{M}$ by 
\begin{equation}
\sigma_{\rm x}: \mathcal{M} \supset U \ni p \mapsto \left\{\frac{\partial}{\partial{x^\mu}}(p)\right\} \in L_p\mathcal{M},
\end{equation}
and conversely, thanks to the Frobenius Theorem, any smooth local section
\begin{equation}
\sigma : \mathcal{M} \supset U \to L\mathcal{M}
\end{equation}
such that $[\sigma_\mu,\sigma_\nu]=0$ provides a local coordinate chart
\begin{equation}
{\rm x}_\sigma: \mathcal{M} \supset U \to V \subset \Reals^n
\end{equation}
via the exponential map; we will restrict to such locally commuting sections being associated with frames, in keeping with their interpretation as arbitrary local coordinate systems. As explained in the previous section, the choice of a metric $g_{\mu\nu}$ on $\mathcal{M}$ amounts to specifying all the local coordinate charts in which $g_{\mu\nu}$ is diagonal, which leaves the $SO(1,3)$-gauge freedom for choosing the tetrad field compatible with the metric. Thus, a metric $g_{\mu\nu}$ can be captured as an \emph{embedding} of a principal $SO(1,3)$-bundle associated to $g_{\mu\nu}$ into the frame bundle of $\mathcal{M}$ \cite{catren_geometrical_2015}
\begin{equation}
    i_{g_{\mu\nu}}: B_{g_{\mu\nu}} \hookrightarrow L\mathcal{M}.
\end{equation}

\subsection{GR-frames}

If we now consider a frame bundle as the principal bundle for our quantum frames, we arrive at the following picture. A QRF on $\pi_L: L\mathcal{M} \to \mathcal{M}$ is a POVM
\begin{equation}
    \E_\R: {\rm Bor}(\pi_L^{-1}(U)) \to \Eff(\hir),
\end{equation}
together with a section $\sigma_L: U \to L\mathcal{M}$ which provides:
\begin{itemize}
    \item a \emph{metric} $g^{\sigma_L}_{\mu\nu}$ on $U$, by declaring $g^\sigma\left(\sigma_\mu,\sigma_\nu\right) := \eta_{\mu\nu}$, 
    \item the \emph{sub-bundle} over $U$ determined by $g^{\sigma_L}_{\mu\nu}$ denoted
    \begin{equation}
         i_{\sigma_L}: B_{g^{\sigma_L}_{\mu\nu}} \hookrightarrow L\mathcal{M},
    \end{equation}

\item a \emph{tetrad field} $\sigma_T: U \to B_{g^{\sigma_L}_{\mu\nu}}$ such that $\sigma_L = i_\sigma \circ \sigma_T$.
\end{itemize}
QRFs on frame bundles will be called \emph{GR-QRFs}. A frame on a curved spacetime $(\mathcal{M},g_{\mu\nu})$ can thus be understood as a \emph{reduction} of a GR-QRF along an embedding $i=i_{g_{\mu\nu}}$.

\subsection{Diffeomorphic frame transformations}

In the context of the frame bundle, a morphism between frames consists of a channel $\psi: B(\hir) \to B(\hirprim)$ compatible with the frame observables as before, and a fiber-preserving map
\begin{equation}
    \theta: \pi_L^{-1}(U) \to \pi_L^{-1}(U')
\end{equation}
such that $\theta \circ \sigma_L = \sigma'_L \circ \varphi_\theta$. Since now the sections $\sigma_L$ provide \emph{arbitrary} local coordinate systems, the maps $\varphi_\theta$ may be given by general (non-isometric) \emph{diffeomorphisms} $U \to U'$. Thus, external frame transformations as defined in this work subsume local coordinate transformations of classical physics.

\subsection{Transcending inertiality}

Although the section of the frame bundle specifies the Lorentzian geometry as in the previous section, now the frames are not bound to be inertial with respect to this geometry, which can be illustrated as follows. Specifying a state $\omega \in \St(\hir)$ of a GR-QRF gives a probability distribution $\mu^{\E_\R}_\omega$ on $\pi_L^{-1}(U)$. If the support of $\mu^{\E_\R}_\omega$ lies in the sub-bundle $B_{g^{\sigma_L}_{\mu\nu}} \subset L\mathcal{M}|_U$ associated to $\sigma_L$, we get
\begin{align} \begin{split}
    \Y^\R_\omega (\hat{\phi}) &= 
    \int_{\pi_L^{-1}(U)} \hat{\phi}(\pi(b)).W_{\sigma_L}(b) \h d\mu^{\E_\R}_\omega(b)\\
    &= \int_{B_{g^{\sigma_L}_{\mu\nu}}} \hat{\phi}(\pi(b)).\Lambda_{\sigma_T}(b) \h d\mu^{\E_\R}_\omega(b),
\end{split} \end{align}
where $W \in GL(4,\Reals)$ and $\Lambda \in SO(1,3)$ as before, recovering the curved space-time picture. The only difference is that now $g_{\mu\nu} = g^{\sigma_L}_{\mu\nu}$ is specified by the frame and not fixed from the outset.

However, $\omega \in \St(\hir)$ may not be compatible with the metric specified by $\sigma_L$, in the sense of the support of $\mu^{\E_\R}_\omega$ lying on some other embedded principal $SO(1,3)$-bundle, or indeed not supported on a single such bundle at all! If this is the case, we may want to interpret $\omega \in \St(\hir)$ as describing a \emph{non-inertial} -- with respect to the $g^{\sigma_L}_{\mu\nu}$ geometry treated as background  -- quantum reference frame. 

\section{QRFs and indefinite geometry}

Here we propose an alternative interpretation of $GR$-QRFs. The section $\sigma_L: U \to L\mathcal{M}$ is needed for the frame to define the relativization for fields, but maybe it does not have to be interpreted as fixing the background geometry. On the contrary, we could insist that the frame is always understood as inertial, in a way by definition, only with the notion of inertiality allowed to vary \emph{probabilistically}. This is the picture we get when interpreting the probabilities
\begin{equation}
    {\rm prob}(g_{\mu\nu},\omega) := \mu^{\E_\R}_\omega(B_{g_{\mu\nu}}) \in [0,1]
\end{equation}
as providing a \emph{probabilistic geometry} for $\mathcal{M}$, conditioned on the quantum state $\omega \in \St(\hir)$. This interpretation could lead to an alternative to other proposals (see e.g. \cite{giacomini_einsteins_2023,giacomini_quantum_2019,oreshkov_quantum_2012}) and mathematically tractable descriptions of quantum systems/fields in \emph{indefinite} (relational) geometry/causal order.

We finish with some speculations concerning the potential usefulness of the introduced tools and concepts in unifying GR with QFT. Considering $SO(1,3)$ acting on $GL(4,\Reals)$ by conjugation, we arrive at a local \emph{stratification} (by orbit types, see e.g. \cite{ross_stratified_2024}) of the frame bundle $L\mathcal{M}$ into a sum of orbits corresponding to different choices of the metric field on $\mathcal{M}$, in that for any small enough $U \subseteq\mathcal{M}$ we have 
\begin{equation}\label{strat2}
    {\rm Frm}(U) \cong U \times {\rm IP}(1,3) \times SO(1,3),
\end{equation}
where ${\rm IP}(1,3):=GL(4,\Reals)/SO(1,3)$ parameterize the pointwise choice of an inner product of Lorentzian signature. Locally we can then write a section of the frame bundle as
\begin{equation}
\sigma (p) = (g^{\sigma_L}_{\mu\nu}(p),\Lambda_{\sigma_T}(p)).
\end{equation}

This perspective suggests viewing the ${\rm IP}(1,3)$ factor in the fibers of the frame bundle as an extra \emph{dynamical} structure: just like we could extend the $SO(1,3)$ principal bundle by additional gauge groups, moving from an $SO(1,3)$-bundle to $L\mathcal{M}$ we are perhaps doing a similar thing, only the new fibers are not homeomorphic to any (gauge) group.

To couple the new, geometric degrees of freedom to the fields we notice that the operator-valued PDE \eqref{sec:PDEs}
\begin{equation}
    \hat{T}: C^\infty(W,B(\his)) \supset \hat{V} \to C^\infty(W,B(\his)),
\end{equation}
that needs to be satisfied by the to-be-relativized quantum fields $\hat{\phi}$, will depend on the metric tensor as it enters the GR action functional from which the Lagrange equations for the fields coupled to geometry are derived. We can incorporate this into the definition of relational operators by writing $\hat{T}[g_{\mu\nu}]$ for the metric-dependent differential operator and putting
\begin{equation}
    \Y^\R (\hat{\phi})_{GR} := \int_{L\mathcal{M}} \delta\left(\hat{T}[g^{\sigma_L}_{\mu\nu}(b)]\hat{\phi}\right)\hat{\phi}(p).\Lambda_{\sigma_T} \otimes d\E_\R(b),
\end{equation}
where $g^{\sigma_L}_{\mu\nu}(b)$ denotes a Lorentzian metric at $\pi(p)$ specified by the basis of $T_{\pi(b)}\mathcal{M}$ arising by transforming $\sigma_L(\pi(b))$ with $W_\sigma(b) \in GL(4,\Reals)$, and 
with $\hat{T}[g_{\mu\nu}]$ understood locally around $p \in \mathcal{M}$. This way we give justice to the fact that the fields should satisfy geometry-dependent equations in the context of indefinite geometries arising in the setup of GR-QRFs.

\section{Summary and outlook}

In this paper, we have proposed a basis on which to build a comprehensive framework for Relational Quantum Field Theory (RQFT) by extending the principles of QRFs within the context of relativistic quantum physics. Our approach integrates the methodology of QRFs into QFT, providing a new operational perspective on the latter, emphasizing the relational nature of spacetime regions and ensuring global gauge invariance while breaking local gauge invariance to obtain relational and operationally meaningful local field observables.

One of the significant contributions of this work is the development of new mathematical tools essential for the formalization and application of QRFs in the context of curved spacetime geometries and beyond. These tools include operator-valued integration and a tentative approach to partial differential equations for operator-valued functions. The former allows for the integration of a wide class of operator-valued functions with respect to arbitrary positive operator-valued measures, providing a foundation for the generalization of the relativization procedure to gauge theories. The latter, although still speculative, offers a new approach to differential equations in the context of quantum fields providing a direct link between the gauge symmetries and differential equations on the quantum level.

The formalism is discussed in the context of curved space-time geometries by employing the formalism of Lorentz bundles. In this case, given a frame and a quantum state provides a probabilistic localization of the frame and orientation of the reference tetrad field. A number of scenarios in which a frame is restricted to follow a curve were described. The notion of a frame transformation in the curved space-time scenarios subsumes that of a space-time isometry.

We have also explored the concept of non-inertial QRFs and the possibility of describing quantum fields on spacetimes without a predetermined metric structure. This generalization allows for the consideration of more general frames, which may not necessarily be inertial with respect to the Lorentzian geometry that they specify. From another perspective, understanding QRFs as inherently inertial, with inertiality allowed to vary probabilistically, leads to a novel perspective on indefinite spacetime geometries arising from quantum states. Finally, we speculate how such indefinite causality could be incorporated into the dynamics by a suitable relativization procedure.

Looking ahead, several pressing developments and intriguing research directions emerge from this work. One immediate area of investigation is the compatibility of our framework with the existing axiomatic approaches to QFT. To this end, relational version of the Wightman axiomatics and the algebraic properties of the algebras generated by relational local quantum observables are being investigated. To fully realize the potential of RQFT, several formal and conceptual advancements are also necessary. These include a deeper exploration of operator-valued integration and partial differential equations for operator-valued functions on manifolds and the properties of field-theoretic relativization.

Concrete models, such as the Klein-Gordon field coupled to a detector modelled as a QRF will be essential in demonstrating the practical applicability of the proposed framework. Providing an operational and relational model capturing a gravitational field sourced by a mass in a superposition of space-time localization is another potential application area of the proposed formalism, that would be of interest to the broad quantum foundations community.

An approach to this last mentioned model-building challenge based on quantum reference frames ideas have been attempted in a recent work \cite{vanzella_frame-bundle_2024}, which seem to come close to the formalism presented here. We will now briefly outline how we see this approach fits into ours.

In \cite{vanzella_frame-bundle_2024}, a quantum reference frame is defined as a complex-valued function on a principal $H$-bundle
\[
    \Psi: B \to \mathbb{C},
\]
where $\pi:B \to \mathcal{M}$, such that for all $p \in \mathcal{M}$
\[
    ||\Psi(p)||^2 := \int_H |\Psi(p,h)|^2d\mu < \infty,
\]
where $\mu$ is the Haar measure on the gauge group $H$, which is taken to be the Lorentz group or the general linear group for the exactly the same reasons that forced us to consider them in this role. If the support $U$ of $\Psi$ is compact, which is assumed in the context of QFT observables, $\Psi$ seem to be nothing else than a non-normalised pure state of an ideal frame, i.e, one modelled on the Hilbert space of the square-intergable functions
\[
    \hir := L^2(U \times H),
\]
with the natural action of $H$, and the covariant frame observable given by
\[
    \E_\R: {\rm Bor}(U \times H) \ni Y \mapsto M_{\chi_Y} \in \Eff(\hir),
\]
where $M_{\chi_Y}$ is the multiplication operator by the characteristic function $\chi_Y$ of the Borel subset $Y$. The function $||\Psi(p)||^2$ is considered to play a smearing role for local quantum field observables, the expectation values of field observables taken relative to quantum frames also seem compatible with the relativisation procedure defined here. The detailed relation between the approaches, with potential cross-fertilisations, will be explored in a separate work.

We also plan to investigate upper bounds on localization of frames in relation to renormalisation, as well as the prospects of intersecting the ideas of indefinite quantum geometries arising in the context of frame bundles with the dynamics of GR using the concepts of spacetime thermodynamics \cite{jacobson_thermodynamics_1995,jacobson_entanglement_2016}.

\paragraph*{Acknowledgments}

I want to thank first Prof. Klaas Landsmann, who encouraged me to study mathematics, and Dr. Leon Loveridge from whom I learned so much.  I am also grateful for a brief but very inspiring conversation with Dr. Yui Kuramochi which encouraged me to pursue my ideas, and Jeremy Butterfield who urged me to write this note. Among other kind listeners of these ideas at various stages of their development I am happy to mention Marius Oancea, Chris Fewster, Nesta van der Shaaf, Anne-Catherine de la Hamette, Eduardo Martin-Martinez, Philipp H\"{o}hn, Aleks Kissinger, Tein van der Lugt and other members of the Quantum Group at Oxford, to whom my gratitude extends.

This publication was made possible through the support of the ID\# 62312 grant from the John Templeton Foundation, as part of the \href{https://www.templeton.org/grant/the-quantuminformation-structure-ofspacetime-qiss-second-phase}{‘The Quantum Information Structure of Spacetime’ Project (QISS)}. The opinions expressed in this project/publication are those of the author(s) and do not necessarily reflect the views of the John Templeton Foundation.

\printbibliography[title={References}]

@book{streater_pct_2000,
	address = {Princeton, N.J},
	edition = {1st pbk. print., with rev. pref. and corrections},
	series = {Princeton landmarks in physics},
	title = {{PCT}, spin and statistics, and all that},
	isbn = {978-0-691-07062-9},
	language = {en},
	publisher = {Princeton University Press},
	author = {Streater, R. F. and Wightman, A. S.},
	year = {2000},
}

@misc{vanzella_frame-bundle_2024,
	title = {A frame-bundle formulation of quantum reference frames: from superposition of perspectives to superposition of geometries},
	shorttitle = {A frame-bundle formulation of quantum reference frames},
	url = {http://arxiv.org/abs/2406.15838},
	doi = {10.48550/arXiv.2406.15838},
	urldate = {2024-07-21},
	publisher = {arXiv},
	author = {Vanzella, Daniel A. Turolla and Butterfield, Jeremy},
	month = jun,
	year = {2024},
	note = {arXiv:2406.15838 [gr-qc, physics:quant-ph]},
}

@unpublished{glowacki_operational_2023,
	title = {Operational {Quantum} {Frames}: {An} operational approach to quantum reference frames},
	shorttitle = {Operational {Quantum} {Frames}},
	url = {http://arxiv.org/abs/2304.07021},
	author = {Głowacki, Jan},
	month = apr,
	year = {2023},
}

@article{glowacki_quantum_2024,
	title = {Quantum {Reference} {Frames} on {Finite} {Homogeneous} {Spaces}},
	volume = {63},
	issn = {1572-9575},
	url = {https://doi.org/10.1007/s10773-024-05650-7},
	doi = {10.1007/s10773-024-05650-7},
	language = {en},
	number = {5},
	urldate = {2024-06-14},
	journal = {International Journal of Theoretical Physics},
	author = {Głowacki, Jan and Loveridge, Leon and Waldron, James},
	month = may,
	year = {2024},
	pages = {137},
}

@article{jacobson_entanglement_2016,
	title = {Entanglement {Equilibrium} and the {Einstein} {Equation}},
	volume = {116},
	issn = {0031-9007, 1079-7114},
	url = {http://arxiv.org/abs/1505.04753},
	doi = {10.1103/PhysRevLett.116.201101},
	number = {20},
	urldate = {2024-05-13},
	journal = {Physical Review Letters},
	author = {Jacobson, Ted},
	month = may,
	year = {2016},
	note = {arXiv:1505.04753 [gr-qc, physics:hep-th]},
	pages = {201101},
}

@article{jacobson_thermodynamics_1995,
	title = {Thermodynamics of {Spacetime}: {The} {Einstein} {Equation} of {State}},
	volume = {75},
	issn = {0031-9007, 1079-7114},
	shorttitle = {Thermodynamics of {Spacetime}},
	url = {https://link.aps.org/doi/10.1103/PhysRevLett.75.1260},
	doi = {10.1103/PhysRevLett.75.1260},
	language = {en},
	number = {7},
	urldate = {2022-12-16},
	journal = {Physical Review Letters},
	author = {Jacobson, Ted},
	month = aug,
	year = {1995},
	pages = {1260--1263},
}

@book{haag_local_1996,
	address = {Berlin, Heidelberg},
	title = {Local {Quantum} {Physics}},
	copyright = {http://www.springer.com/tdm},
	isbn = {978-3-540-61049-6 978-3-642-61458-3},
	url = {http://link.springer.com/10.1007/978-3-642-61458-3},
	urldate = {2024-05-24},
	publisher = {Springer},
	author = {Haag, Rudolf},
	year = {1996},
	doi = {10.1007/978-3-642-61458-3},
}

@article{perche_particle_2024,
	title = {Particle detectors from localized quantum field theories},
	volume = {109},
	url = {https://link.aps.org/doi/10.1103/PhysRevD.109.045013},
	doi = {10.1103/PhysRevD.109.045013},
	number = {4},
	urldate = {2024-05-23},
	journal = {Physical Review D},
	author = {Perche, T. Rick and Polo-Gómez, José and Torres, Bruno de S. L. and Martín-Martínez, Eduardo},
	month = feb,
	year = {2024},
	note = {Publisher: American Physical Society},
	pages = {045013},
}

@article{fraser_note_2023,
	title = {Note on episodes in the history of modeling measurements in local spacetime regions using {QFT}},
	volume = {48},
	issn = {2102-6467},
	url = {https://doi.org/10.1140/epjh/s13129-023-00064-1},
	doi = {10.1140/epjh/s13129-023-00064-1},
	language = {en},
	number = {1},
	urldate = {2024-05-23},
	journal = {The European Physical Journal H},
	author = {Fraser, Doreen and Papageorgiou, Maria},
	month = nov,
	year = {2023},
	pages = {14},
}

@article{polo-gomez_detector-based_2022,
	title = {A detector-based measurement theory for quantum field theory},
	volume = {105},
	issn = {2470-0010, 2470-0029},
	url = {http://arxiv.org/abs/2108.02793},
	doi = {10.1103/PhysRevD.105.065003},
	number = {6},
	urldate = {2024-05-23},
	journal = {Physical Review D},
	author = {Polo-Gómez, José and Garay, Luis J. and Martín-Martínez, Eduardo},
	month = mar,
	year = {2022},
	note = {arXiv:2108.02793 [gr-qc, physics:hep-th, physics:quant-ph]},
	pages = {065003},
}

@misc{carette_operational_2023,
	title = {Operational {Quantum} {Reference} {Frame} {Transformations}},
	url = {http://arxiv.org/abs/2303.14002},
	doi = {10.48550/arXiv.2303.14002},
	urldate = {2024-05-22},
	publisher = {arXiv},
	author = {Carette, Titouan and Głowacki, Jan and Loveridge, Leon},
	month = dec,
	year = {2023},
	note = {arXiv:2303.14002 [math-ph, physics:quant-ph]},
}

@article{glowacki_operator-valued_2024,
	title = {Operator-valued {Integration}},
	journal = {In preparation},
	author = {Głowacki, Jan},
	month = may,
	year = {2024},
}

@misc{ross_stratified_2024,
	title = {Stratified {Vector} {Bundles}: {Examples} and {Constructions}},
	shorttitle = {Stratified {Vector} {Bundles}},
	url = {http://arxiv.org/abs/2303.04200},
	doi = {10.48550/arXiv.2303.04200},
	urldate = {2024-05-17},
	publisher = {arXiv},
	author = {Ross, Ethan},
	month = jan,
	year = {2024},
	note = {arXiv:2303.04200 [math]},
}

@article{oreshkov_quantum_2012,
	title = {Quantum correlations with no causal order},
	volume = {3},
	copyright = {2012 The Author(s)},
	issn = {2041-1723},
	url = {https://www.nature.com/articles/ncomms2076},
	doi = {10.1038/ncomms2076},
	language = {en},
	number = {1},
	urldate = {2024-05-17},
	journal = {Nature Communications},
	author = {Oreshkov, Ognyan and Costa, Fabio and Brukner, Časlav},
	month = oct,
	year = {2012},
	note = {Publisher: Nature Publishing Group},
	pages = {1092},
}

@misc{chen_quantum_2024,
	title = {Quantum effects in gravity beyond the {Newton} potential from a delocalised quantum source},
	url = {http://arxiv.org/abs/2402.10288},
	urldate = {2024-05-17},
	publisher = {arXiv},
	author = {Chen, Lin-Qing and Giacomini, Flaminia},
	month = feb,
	year = {2024},
	note = {arXiv:2402.10288 [gr-qc, physics:quant-ph]},
}

@incollection{rejzner_algebraic_2016,
	address = {Cham},
	title = {Algebraic {Approach} to {Quantum} {Theory}},
	isbn = {978-3-319-25901-7},
	url = {https://doi.org/10.1007/978-3-319-25901-7_2},
	language = {en},
	urldate = {2024-05-17},
	booktitle = {Perturbative {Algebraic} {Quantum} {Field} {Theory}: {An} {Introduction} for {Mathematicians}},
	publisher = {Springer International Publishing},
	author = {Rejzner, Kasia},
	editor = {Rejzner, Kasia},
	year = {2016},
	doi = {10.1007/978-3-319-25901-7_2},
	pages = {3--37},
}

@misc{fewster_quantum_2024,
	title = {Quantum reference frames, measurement schemes and the type of local algebras in quantum field theory},
	url = {http://arxiv.org/abs/2403.11973},
	doi = {10.48550/arXiv.2403.11973},
	urldate = {2024-05-16},
	publisher = {arXiv},
	author = {Fewster, Christopher J. and Janssen, Daan W. and Loveridge, Leon Deryck and Rejzner, Kasia and Waldron, James},
	month = mar,
	year = {2024},
	note = {arXiv:2403.11973 [gr-qc, physics:hep-th, physics:math-ph, physics:quant-ph]},
}

@article{mazzucchi_observables_2001,
	title = {On the observables describing a quantum reference frame},
	volume = {42},
	issn = {0022-2488},
	url = {https://doi.org/10.1063/1.1370395},
	doi = {10.1063/1.1370395},
	number = {6},
	urldate = {2024-04-24},
	journal = {Journal of Mathematical Physics},
	author = {Mazzucchi, S.},
	month = jun,
	year = {2001},
	pages = {2477--2489},
}

@misc{glowacki_relativization_2024,
	title = {Relativization is naturally functorial},
	url = {http://arxiv.org/abs/2403.03755},
	doi = {10.48550/arXiv.2403.03755},
	publisher = {arXiv},
	author = {Głowacki, Jan},
	month = mar,
	year = {2024},
}

@article{catren_geometrical_2015,
	title = {Geometrical {Foundations} of {Cartan} {Gauge} {Gravity}},
	volume = {12},
	issn = {0219-8878, 1793-6977},
	url = {http://arxiv.org/abs/1407.7814},
	doi = {10.1142/S0219887815300020},
	number = {04},
	urldate = {2024-04-02},
	journal = {International Journal of Geometric Methods in Modern Physics},
	author = {Catren, Gabriel},
	month = apr,
	year = {2015},
	note = {arXiv:1407.7814 [gr-qc]},
	pages = {1530002},
}

@misc{kabel_identification_2024,
	title = {Identification is {Pointless}: {Quantum} {Reference} {Frames}, {Localisation} of {Events}, and the {Quantum} {Hole} {Argument}},
	shorttitle = {Identification is {Pointless}},
	url = {http://arxiv.org/abs/2402.10267},
	urldate = {2024-02-26},
	publisher = {arXiv},
	author = {Kabel, Viktoria and de la Hamette, Anne-Catherine and Apadula, Luca and Cepollaro, Carlo and Gomes, Henrique and Butterfield, Jeremy and Brukner, Časlav},
	month = feb,
	year = {2024},
	note = {arXiv:2402.10267 [gr-qc, physics:physics, physics:quant-ph]},
}

@article{giacomini_spacetime_2021,
	title = {Spacetime {Quantum} {Reference} {Frames} and superpositions of proper times},
	volume = {5},
	url = {https://quantum-journal.org/papers/q-2021-07-22-508/},
	urldate = {2024-02-21},
	journal = {Quantum},
	author = {Giacomini, Flaminia},
	year = {2021},
	note = {Publisher: Verein zur Förderung des Open Access Publizierens in den Quantenwissenschaften},
	pages = {508},
}

@article{giacomini_quantum_2019,
	title = {Quantum mechanics and the covariance of physical laws in quantum reference frames},
	volume = {10},
	url = {https://www.nature.com/articles/s41467-018-08155-0},
	number = {1},
	urldate = {2024-02-21},
	journal = {Nature communications},
	author = {Giacomini, Flaminia and Castro-Ruiz, Esteban and Brukner, Časlav},
	year = {2019},
	note = {Publisher: Nature Publishing Group UK London},
	pages = {494},
}

@misc{giacomini_einsteins_2023,
	title = {Einstein's {Equivalence} principle for superpositions of gravitational fields and quantum reference frames},
	url = {http://arxiv.org/abs/2012.13754},
	urldate = {2024-02-19},
	publisher = {arXiv},
	author = {Giacomini, Flaminia and Brukner, Časlav},
	month = jun,
	year = {2023},
	note = {arXiv:2012.13754 [gr-qc, physics:quant-ph]},
}

@book{busch_quantum_1996,
	address = {Berlin Heidelberg},
	edition = {2., rev. ed},
	series = {Lecture notes in physics {New} series {M}, monographs},
	title = {The quantum theory of measurement},
	isbn = {978-3-540-61355-8},
	language = {en},
	number = {2},
	publisher = {Springer},
	author = {Busch, Paul and Lahti, Pekka J. and Mittelstaedt, Peter and Lahti, Pekka Johannes},
	year = {1996},
}

@misc{fewster_measurement_2023,
	title = {Measurement in {Quantum} {Field} {Theory}},
	url = {http://arxiv.org/abs/2304.13356},
	urldate = {2023-10-25},
	publisher = {arXiv},
	author = {Fewster, Christopher J. and Verch, Rainer},
	month = apr,
	year = {2023},
	note = {arXiv:2304.13356 [gr-qc, physics:hep-th, physics:math-ph]},
}

@article{loveridge_relative_2019,
	title = {Relative {Quantum} {Time}},
	volume = {49},
	issn = {1572-9516},
	url = {https://doi.org/10.1007/s10701-019-00268-w},
	doi = {10.1007/s10701-019-00268-w},
	language = {en},
	number = {6},
	urldate = {2023-10-23},
	journal = {Foundations of Physics},
	author = {Loveridge, Leon and Miyadera, Takayuki},
	month = jun,
	year = {2019},
	pages = {549--560},
}

@phdthesis{loveridge_quantum_2012,
	type = {phd},
	title = {Quantum {Measurements} in the {Presence} of {Symmetry}},
	copyright = {cc\_by\_nc\_nd},
	url = {https://etheses.whiterose.ac.uk/2670/},
	language = {en},
	urldate = {2023-10-23},
	school = {University of York},
	author = {Loveridge, Leon},
	month = jul,
	year = {2012},
}

@article{loveridge_relativity_2017,
	title = {Relativity of quantum states and observables},
	volume = {117},
	url = {https://iopscience.iop.org/article/10.1209/0295-5075/117/40004/meta},
	number = {4},
	urldate = {2023-10-23},
	journal = {Europhysics Letters},
	author = {Loveridge, Leon and Busch, Paul and Miyadera, Takayuki},
	year = {2017},
	note = {Publisher: IOP Publishing},
	pages = {40004},
}

@article{loveridge_symmetry_2018,
	title = {Symmetry, {Reference} {Frames}, and {Relational} {Quantities} in {Quantum} {Mechanics}},
	volume = {48},
	issn = {1572-9516},
	url = {https://doi.org/10.1007/s10701-018-0138-3},
	doi = {10.1007/s10701-018-0138-3},
	language = {en},
	number = {2},
	urldate = {2023-10-23},
	journal = {Foundations of Physics},
	author = {Loveridge, Leon and Miyadera, Takayuki and Busch, Paul},
	month = feb,
	year = {2018},
	pages = {135--198},
}

@article{ali_systems_1998,
	title = {Systems of {Covariance} in {Relativistic} {Quantum} {Mechanics}},
	volume = {37},
	issn = {1572-9575},
	url = {https://doi.org/10.1023/A:1026687306214},
	doi = {10.1023/A:1026687306214},
	language = {en},
	number = {1},
	urldate = {2023-10-17},
	journal = {International Journal of Theoretical Physics},
	author = {Ali, S. Twareque},
	month = jan,
	year = {1998},
	pages = {365--373},
}

@book{takesaki2001theory,
	series = {Encyclopaedia of mathematical sciences},
	title = {Theory of operator algebras {I}},
	isbn = {978-3-540-42248-8},
	url = {https://books.google.pl/books?id=dTnq4hjjtgMC},
	publisher = {Springer Berlin Heidelberg},
	author = {Takesaki, M.},
	year = {2001},
	note = {tex.lccn: 79013655},
}

@article{loveridge_relational_2020,
	title = {A relational perspective on the {Wigner}-{Araki}-{Yanase} theorem},
	volume = {1638},
	issn = {1742-6588, 1742-6596},
	url = {http://arxiv.org/abs/2006.07047},
	doi = {10.1088/1742-6596/1638/1/012009},
	number = {1},
	urldate = {2023-09-04},
	journal = {Journal of Physics: Conference Series},
	author = {Loveridge, Leon},
	month = oct,
	year = {2020},
	note = {arXiv:2006.07047 [quant-ph]},
	pages = {012009},
}

@misc{de_la_hamette_perspective-neutral_2021,
	title = {Perspective-neutral approach to quantum frame covariance for general symmetry groups},
	url = {http://arxiv.org/abs/2110.13824},
	doi = {10.48550/arXiv.2110.13824},
	urldate = {2023-08-30},
	publisher = {arXiv},
	author = {de la Hamette, Anne-Catherine and Galley, Thomas D. and Hoehn, Philipp A. and Loveridge, Leon and Mueller, Markus P.},
	month = oct,
	year = {2021},
	note = {arXiv:2110.13824 [gr-qc, physics:hep-th, physics:quant-ph]},
}

\appendix

\section{POVMs}\label{POVMs}

A positive operator-valued measure is a direct analogue of a probability measure.\footnote{We note that some authors do not assume normalization of POVMs, as we do.} Given a measurable space $(\Sigma,\F)$, so a set $\Sigma$ and a family ($\sigma$-algebra) $\F$ of subsets of $\Sigma$, a POVM on $(\Sigma,\F)$ is a set function with values in the operator algebra $B(\hi)$ on a Hilbert space $\hi$, i.e,
\[
    \E_\R: \F \to B(\hi),
\]
such that for any $\omega \in \S(\hi)$ the associated set function
\[
    \mu^{\E_\R}_\omega\F \ni X \longmapsto  \tr[\omega \E_\R(X)] \in [0,1]
\]
is a \emph{probability measure}. This can only happen if for all $X \in \F$ we have ${\E_\R}(X) \in \Eff(\hi)$, so that in fact we have an \emph{effect-valued measure}, with the operators ${\E_\R}(X)$ are rightly called the \emph{effects of} ${\E_\R}$. POVMs then assign probability measures to quantum states
\[
\S(\hi) \ni \omega \mapsto \mu^{\E_\R}_\omega \in {\rm Prob}(\Sigma,\F).
\]
In a sense, this is the most general way this can be achieved as any assignment
\[
\S(\hi) \ni \omega \mapsto \mu_\omega \in {\rm Prob}(\Sigma,\F)
\]
such that for any $X \in \F$ the map $\omega \mapsto \mu_\omega(X)$ is (trace-norm) continuous, needs to be given via~a~POVM.

\section{Proof of Thm. \ref{thm:intop}}\label{app:proof}

Here we give a proof of Thm. \ref{thm:intop}.

\begin{proof}
We will show that the map assigning to product states on $B(\hisr)$ the integrals \eqref{integrals} uniquely extends to a bounded linear operator in $B(\hisr)$. Let us first check how $(\omega, \rho) \mapsto \int_\Sigma \tr[\rho \h f(x)] d\mu^{\E_\R}_\omega$ behaves with respect to the vector space structures on $\T(\hir)$ and $\T(\his)$. To this end, notice that for any $\lambda \in \Cn$, $\omega \in \S(\hir)$ and $\rho \in \S(\his)$ we have
    \begin{align} \begin{split}
        &\lambda \int_\Sigma \tr[\rho \h f(x)] d\mu^{\E_\R}_\omega = \\
         &\int_\Sigma \tr[(\lambda \rho) f(x)] d\mu^{\E_\R}_\omega =
         \int_\Sigma \tr[\rho \h f(x)] d\mu^{\E_\R}_{\lambda \omega},
    \end{split} \end{align}
    as easily follows from linearity of the trace since $\mu_{\lambda \omega}^{\E_\R}(X) = \tr[(\lambda \omega) \E_\R(X)]$. Similarly, we get
    \begin{align} \begin{split}
        &\int_\Sigma \tr[\rho \h f(x)] d\mu^{\E_\R}_{(\omega + \omega')} =\\
        &\int_\Sigma \tr[\rho \h f(x)] d\mu^{\E_\R}_{\omega} +
        \int_\Sigma \tr[\rho \h f(x)] d\mu^{\E_\R}_{\omega'},
    \end{split} \end{align}
    while the linearity of the Lebesgue integration gives
    \begin{align} \begin{split}
        &\int_\Sigma \tr[(\rho + \rho') f(x)] d\mu^{\E_\R}_\omega =\\
        &\int_\Sigma \tr[\rho \h f(x)] d\mu^{\E_\R}_\omega + \int_\Sigma \tr[\rho' f(x)] d\mu^{\E_\R}_\omega.
    \end{split} \end{align}
    The assignment $(\omega, \rho) \mapsto \int_\Sigma \tr[\rho \h f(x)] d\mu^{\E_\R}_\omega$ can thus be uniquely extended to a \emph{bilinear} functional
    \begin{equation}
    \phi_{(f,\E_\R)}: \T(\hir) \times \T(\his) \to \Cn.
    \end{equation}
Notice here that $\phi_{(f,\E_\R)}$ is bounded if and only if all the integrals \eqref{integrals} converge, which is the case when $f$ is ultraweakly continuous and bounded since then all $f_\rho$ are continuous and bounded and hence integrable with respect to finite measures, including $\mu^{\E_\R}_\omega$ for all $\omega$.

Now the universal property of the tensor product \emph{of vector spaces} gives the unique linear map
\begin{equation}
\tilde{\phi}_{(f,\E_\R)}: \T(\his) \otimes \T(\hir) \to \Cn.
\end{equation}
Denote the restriction of $\tilde{\phi}_{(f,\E_\R)}$ to the product states $\S(\hisr)_{\rm prod} \subset \T(\hisr)$ by $\phi_{A_{(f,\E_\R)}}$. This map is \emph{affine} by definition and bounded since
\begin{equation}
        ||\tilde{\phi}_{(f,\E_\R)}||_\infty
        = \sup_{\omega, \rho} \left| \int_\Sigma \tr[\rho \h f(x)] d\mu^{\E_\R}_\omega(x) \right|< \infty,
\end{equation}
where the supremum is taken over $\omega \in \S(\hir)$, $\rho \in \S(\his)$. Now since any bounded linear operator $A \in B(\hi)$ is uniquely specified by a bounded affine map \cite{takesaki2001theory}
\begin{equation}
            \phi_A: \S(\hi) \ni \rho \mapsto \tr[\rho \h A] \in \Cn,
\end{equation} 
and the map $\phi_{A_{(f,\E_\R)}}$ uniquely extends to a bounded linear functional on the whole of $\T(\hisr)$, it indeed identifies a unique bounded linear operator $A_{(f,\E_\R)} \in B(\hisr)$ satisfying \eqref{integrals}.

The behaviour of the $A_{(f,\E_\R)}$ operators with respect to pull-back and push-forward maps given by measurable functions $\varphi: (\Sigma,\F) \to (\Sigma',\F')$ follows straight from the properties of the Lebesgue integration. Indeed, for any $\rho \in \S(\his), \omega \in \S(\hir)$ we have
    \begin{align} \begin{split}
        &\tr[\rho \otimes \omega \h A_{(f',\E_\R \circ \varphi^{-1})}] = \int_{\Sigma} f'_\rho \h d(\mu^{\E_\R}_\omega \circ \varphi^{-1}) =\\
        & \int_{\Sigma} f'_\rho \circ \varphi \h d\mu^{\E_\R}_\omega =
        \tr[\rho \otimes \omega \h A_{(f' \circ \varphi, \E_\R)}],
    \end{split} \end{align}
and since the $A_{(f,\E_\R)}$ operators are completely determined by their values as functionals on product states, this gives the claim. The last property follows similarly as we have
\begin{align*}
    &\tr[\rho \otimes \omega \h A_{(f,\psi \circ \E_\R)}]  
    = \int_{\Sigma} f_\rho \h d\mu^{\psi \circ {\E_\R}}_\omega
    = \int_{\Sigma} f_\rho \h d\mu^{\E_\R}_{\psi_*(\omega)}\\
    &= \tr[\rho \otimes \psi_*(\omega) \h A_{(f,\E_\R)}]
    = \tr[\rho \otimes \omega \h \id_{\his} \otimes \psi \left(A_{(f,\E_\R)}\right)],
\end{align*}
where the second equality holds since
\begin{align*}
\mu^{\psi \circ {\E_\R}}_\omega(X) &= \tr[\omega \h \psi(\E_\R(X))]\\
&= \tr[\psi_*(\omega) \h \E_\R(X)] = \mu^{\E_\R}_{\psi_*(\omega)}(X).
\end{align*}
Setting $\int_\Sigma f(x) \otimes d\E_\R(x) := A_{(f,\E_\R)}$ finishes the proof.
\end{proof}
\end{multicols}
\end{document}